\documentclass{aa}  

\newcommand*{\ct}{\ensuremath{C_{T}\ }}
\newcommand*{\rt}{\ensuremath{r_{T}^{50}\ }}

\newcommand*{\sigmasfr}{\ensuremath{\Sigma_{SFR}\ }}
\newcommand*{\vism}{\ensuremath{v_\text{IS}\ }}
\newcommand*{\vismFeII}{\ensuremath{v_\text{IS,Fe\,{\scshape ii}}\ }}
\newcommand*{\vismAlIII}{\ensuremath{v_\text{IS,Al\,{\scshape iii}}\ }}

\newcommand*{\vismSiIV}{\ensuremath{v_\text{IS,Si\,{\scshape iv}}\ }}
\newcommand*{\vmaxSiIV}{\ensuremath{v_\text{max,Si\,{\scshape iv}}\ }}

\newcommand*{\vismSiIIb}{\ensuremath{v_\text{IS,Si\,{\scshape ii}}\ }}
\newcommand*{\vismcomb}{\ensuremath{v_\text{IS,comb}\ }}
\newcommand*{\vmaxcomb}{\ensuremath{v_\text{max,comb}\ }}
\newcommand*{\mass}{\ensuremath{M_\star\ }}
\newcommand*{\msun}{\ensuremath{M_\odot\ }}

\newcommand*{\zsys}{\ensuremath{z_{sys}\ }}
\newcommand*{\vmax}{\ensuremath{v_{max}\ }}
\newcommand*{\Mout}{\ensuremath{\dot{M}_{out}\ }}

\newcommand*{\xSiIIa}{Si\,{\scshape ii}\,\ensuremath{\lambda1260\ }}
\newcommand*{\xSiIIb}{Si\,{\scshape ii}\,\ensuremath{\lambda1526\ }}
\newcommand*{\xFeII}{Fe\,{\scshape ii}\,\ensuremath{\lambda1608\ }}
\newcommand*{\xCII}{C\,{\scshape ii}\,\ensuremath{\lambda1334\ }}
\newcommand*{\xOISiII}{O\,{\scshape i}\,Si\,{\scshape ii}\,\ensuremath{\lambda\lambda1302\text{-}1304\ }}
\newcommand*{\xAlII}{Al\,{\scshape ii}\,\ensuremath{\lambda1670\ }}
\newcommand*{\xAlIII}{Al\,{\scshape iii}\,\ensuremath{\lambda\lambda1854\text{-}1862\ }}
\newcommand*{\xSiIV}{Si\,{\scshape iv}\,\ensuremath{\lambda\lambda1393\text{-}1402\ }}
\newcommand*{\xCIII}{C\,{\scshape iii}\,\ensuremath{\lambda\lambda1907\text{-}1909\ }}
\newcommand*{\xCIV}{C\,{\scshape iv}\,\ensuremath{\lambda\lambda1548\text{-}1550\ }}
\newcommand*{\xHeII}{He\,{\scshape ii}\,\ensuremath{\lambda1640\ }}

\newcommand*{\xOIII}{O\,{\scshape iii}\,\ensuremath{\lambda1666\ }}

\newcommand*{\SiII}{Si\,{\scshape ii}\ }
\newcommand*{\SiIIa}{Si\,{\scshape ii}a\ }
\newcommand*{\SiIIb}{Si\,{\scshape ii}b\ }
\newcommand*{\FeII}{Fe\,{\scshape ii}\ }
\newcommand*{\CII}{C\,{\scshape ii}\ }
\newcommand*{\OISiII}{O\,{\scshape i}\,Si\,{\scshape ii}\ }
\newcommand*{\AlII}{Al\,{\scshape ii}\ }
\newcommand*{\AlIII}{Al\,{\scshape iii}\ }
\newcommand*{\SiIV}{Si\,{\scshape iv}\ }
\newcommand*{\CIII}{C\,{\scshape iii]}\ }
\newcommand*{\NV}{N\,{\scshape v}\ }
\newcommand*{\CIV}{C\,{\scshape iv}\ }
\newcommand*{\HeII}{He\,{\scshape ii}\ }

\newcommand*{\OII}{O\,{\scshape ii}\ }

\usepackage{graphicx}

\usepackage[colorlinks=true,
            linkcolor=red,
            urlcolor=green,
            citecolor=blue]{hyperref}

\usepackage{txfonts}

\newcommand{\myemail}{antonello.calabro@inaf.it}

\usepackage{multirow,array}
\usepackage[normalem]{ulem}

\begin{document}



\title{Properties of the interstellar medium in star-forming galaxies at redshifts $2\leq z \leq 5$ from the VANDELS survey}

\author{A. Calabr{\`o}\inst{1} 
\and L. Pentericci\inst{1} 
\and M. Talia\inst{2,3}
\and G. Cresci\inst{4}
\and M. Castellano\inst{1}
\and D. Belfiori\inst{1}
\and S. Mascia\inst{1}
\and G. Zamorani\inst{2}
\and\\ R. Amor\'in\inst{5,6}
\and J.P.U. Fynbo\inst{7} 
\and M. Ginolfi\inst{8}
\and L. Guaita\inst{9}
\and N.P. Hathi\inst{10} 
\and A. Koekemoer\inst{10} 
\and M. Llerena\inst{6} 
\and\\ F. Mannucci\inst{4}
\and P. Santini\inst{1}
\and A. Saxena\inst{11}
\and D. Schaerer\inst{12}
}


\institute{INAF - Osservatorio Astronomico di Roma, via di Frascati 33, 00078, Monte Porzio Catone, Italy (\myemail)
\and INAF - Osservatorio Astronomico di Bologna, via P. Gobetti 93/3, I-40129, Bologna, Italy
\and University of Bologna - Department of Physics and Astronomy `Augusto Righi' (DIFA), Via Gobetti 93/2, I-40129, Bologna, Italy
\and INAF - Osservatorio Astrofisico di Arcetri, Largo E. Fermi 5, I-50125, Firenze, Italy
\and Instituto de Investigaci\'on Multidisciplinar en Ciencia y Tecnolog\'ia, Universidad de La Serena, Ra\'ul Bitr\'an 1305, La Serena, Chile
\and Departamento de F\'isica y Astronom\'ia, Universidad de La Serena, Av. Juan Cisternas 1200 Norte, La Serena, Chile
\and Cosmic Dawn Center (DAWN), Niels Bohr Institute, University of Copenhagen, Jagtvej 128, DK-2200 Copenhagen \O, Denmark
\and European Southern Observatory, Karl Schwarzschild Stra{\ss}e 2, 85748 Garching, Germany
\and Departamento de Ciencias Fisicas, Universidad Andres Bello, Fernandez Concha 700, Las Condes, Santiago, Chile
\and Space Telescope Science Institute, 3700 San Martin Drive, Baltimore, MD 21218, USA
\and Department of Physics and Astronomy, University College London, Gower Street, London WC1E 6BT, UK
\and Department of Astronomy, University of Geneva, 51 Chemin Pegasi, 1290 Versoix, Switzerland
}
\date{Submitted to A\&A}

\abstract 
{
Gaseous flows inside and outside galaxies are key to understanding galaxy evolution, as they regulate their star formation activity and chemical enrichment across cosmic time. 
We study the interstellar medium (ISM) kinematics of a sample of $330$ galaxies with \CIII or \HeII emission using far-ultraviolet (far-UV) ISM absorption lines detected in the ultra deep spectra of the VANDELS survey. 
These galaxies span a broad range of stellar masses from $10^8$ to $10^{11}$ \msun, and star formation rates (SFRs) from $1$ to $500$ \msun/yr in the redshift range between $2$ and $5$. 
We find that the bulk ISM velocity along the line of sight (\vism) is globally in outflow, with a \vism of $-60 \pm 10$ km/s for low-ionisation gas traced by Si\,{\scshape ii}\,\ensuremath{\lambda1260\ \AA }, C\,{\scshape ii}\,\ensuremath{\lambda1334\ \AA}, Si\,{\scshape ii}\,\ensuremath{\lambda1526\ \AA}, and Al\,{\scshape ii}\,\ensuremath{\lambda1670\ \AA} absorption lines, and a \vism of $-160\pm30$ and $-170\pm30$ km/s for higher ionisation gas traced respectively by Al\,{\scshape iii}\,\ensuremath{\lambda\lambda1854\text{-}1862\ \AA} and Si\,{\scshape iv}\,\ensuremath{\lambda\lambda1393\text{-}1402\ \AA}. Interestingly, we notice that BPASS models are able to better reproduce the stellar continuum around the \SiIV doublet than other stellar population templates.  
For individual galaxies, $34\%$ of the sample has a positive ISM velocity shift, almost double the fraction reported at lower redshifts. We additionally derive a maximum outflow velocity \vmax for the average population, which is of the order of $\sim -500$ and $\sim -600$ km/s for the lower and higher ionisation lines, respectively.
Comparing \vism to the host galaxies properties, we find no significant correlations with stellar mass \mass or SFR, and only a marginally significant dependence (at $\sim 2\sigma$) on morphology-related parameters, with slightly higher velocities found in galaxies of smaller size (probed by the equivalent radius \rt), higher concentration (\ct), and higher SFR surface density \sigmasfr. 
From the spectral stacks, \vmax shows a similarly weak dependence on physical properties (at $\simeq 2 \sigma$). 
Moreover, we do not find evidence of enhanced outflow velocities in visually identified mergers compared to isolated galaxies. 
From a physical point of view, the outflow properties are consistent with accelerating momentum-driven winds, with densities decreasing towards the outskirts.
Our moderately lower ISM velocities compared to those found in similar studies at lower redshifts suggest that inflows and internal turbulence might play an increased role at $z>2$ and weaken the outflow signatures. Finally, we estimate mass-outflow rates \Mout that are comparable to the SFRs of the galaxies (hence a mass-loading factor $\eta$ of the order of unity), and an average escape velocity of $625$ km/s, suggesting that most of the ISM will remain bound to the galaxy halo. 
} 


\titlerunning{\footnotesize ISM kinematics from VANDELS far-UV spectra}
\authorrunning{A.Calabr\`o et al.}
 \maketitle
\section{Introduction}\label{introduction}

Galactic inflows and outflows are the main actors of the baryon cycle inside and outside galaxies, playing a fundamental role in the regulation of galaxy evolution across cosmic time. These phenomena of gas flows of the interstellar and circumgalactic medium (ISM and CGM, respectively) are thought to be essential for explaining the discrepancy at low and high masses between the observed shape of the galaxy luminosity function and the predicted mass function of dark matter haloes \citep{madau96,behroozi13}. In addition, they are fundamental ingredients in the explanation of other important scaling relations, including the mass--metallicity relation \citep[MZR,][]{mannucci09,dave11,calabro17,fontanot21} and the star formation rate (SFR)--stellar mass (M$_\star$) relation \citep[e.g.][]{lilly13,tacchella16,rodriguezpuebla16}. 

At lower stellar masses (\mass $\lesssim 2 \times 10^{10}$ \msun), the shape of the luminosity function can be reproduced by considering outflows driven by supernova explosions, stellar winds from OB and Wolf-Rayet stars, UV radiation pressure, and cosmic rays \citep{chevalier77,veilleux05,murray05,hopkins14,fontanot17}. The energy and momentum transferred to the ISM is able to expel part of the gas outside of their shallow potential wells with typical velocities of a few hundred kilometres per second \citep{chevalier85,shapley03}, thus removing the fuel for further star formation. 

When we move to higher halo masses, the above processes are typically not sufficient for the bulk of the gas to reach the velocities needed to escape from the galaxy potential well. 
In these cases, the energetic feedback from an active galactic nucleus (AGN) is thought to be the main factor responsible for the low efficiency in the conversion of baryons into stars \citep[see the review by][]{harrison17}. In this regime, AGN feedback can deposit energy in the surrounding ISM and lead to the ejection of large-scale, high-velocity, and massive winds \citep{fabian12,harrison12,concas22}, finally interrupting the star-formation activity in the host galaxy and the growth of the central supermassive black hole \citep{delucia06,croton06,cattaneo09,kormendy13,forsterschreiber19}. 

In addition to the above-mentioned outflows, gaseous material should also be moving inwards, providing the fuel for star formation and for the build up of disc galaxies \citep{dekel06,dekel09,silk12}. 
These inflows can originate from the condensation of metal-enriched gas that was deposited in a hot corona around galaxies by stellar and AGN winds from previous star-formation episodes. In this case, the infall of gas is part of a circular process that is called the galactic fountain, a phenomenon that is able to sustain the star formation activity of a galaxy  for a long time \citep{marasco12,fraternali17}.
Numerical simulations also predict the infall of more metal-poor and cold gas from the cosmic web over a dynamical timescale that depends on both redshift and host halo mass. At redshifts $\leq 2$, this cold flow of accretion towards the galactic disc is efficient only for lower mass halos ($\lesssim 6 \times 10^{11}$ \msun). For more massive halos, the intracluster medium (ICM) is shock heated at their virial radii to temperatures of $10^6$- $10^7$ K during gravitational collapse, preventing cold gas from penetrating \citep{dekel06}.
At higher redshifts, cold and dense gaseous flows can reach galaxies without strong shocks, leading the cold mode accretion to dominate galaxy growth at these epochs \citep{katz03,keres05,ocvirk08,brooks09}. Promising observational evidence has been found in recent years of cold gas accretion through filamentary structures feeding galaxies in massive halos  \citep[e.g.][]{cantalupo14,hennawi15,daddi21}. 
Overall, galaxies tend to evolve towards a nearly stationary state, where inflow and outflow rates balance their SFR and determine the level of metallicity at fixed stellar mass \citep{bouche10,schaye10,dave12,lilly13,dekel13,bothwell13}. 

In addition to smooth and continuous gas flows from the cosmic web or from the galactic reservoirs, interactions and mergers are able to convey large quantities of gas towards the galaxy centre in a relatively short time ($100$ Myr - $1$ Gyr, depending on the impact parameter, mass ratio, and orientation), triggering intense star-formation episodes, extreme cases of which are called starbursts \citep{rodighiero11,calabro17,calabro18}. Even though mergers are less important in the growth of galaxies than direct cosmological accretion by about an order of magnitude \citep{lhuillier12,combes13}, their contribution to the mass growth rises at earlier cosmic times proportionally to ($1+z$)$^\gamma$, with $\gamma=2.2$-$2.5$ \citep{dayal18}.

Gas flows are detected in galaxies through absorption and emission lines across all ranges of the electromagnetic spectrum, with their typical signatures being a broad wing (often asymmetric) on top of a narrow absorption or emission line component, or an ISM absorption profile whose peak is displaced by several hundreds of kilometres per second compared to emission lines tracing the bulk of the stellar emission. 
Optical and UV spectroscopic surveys have so far detected and characterised ISM velocities in systems ranging from normal star forming galaxies to more extreme starburst and infrared-luminous galaxies in the local Universe \citep[e.g.][]{chisholm15,heckman15}, 
at $1<z<2$ \citep{erb12,rubin14}, and at $z\geq 2$ \citep{pettini02,steidel10}.
All these works agree on the fact that ISM outflows are ubiquitous in absorption at any cosmic epoch and detected in multiple gas phases, with average outflow velocities ranging from $\sim100$ to $\sim200$ km/s (see also \citet{veilleux20} for a review). 
Moreover, evidence of gas inflows from redshifted absorption lines have also been reported in the literature for galaxies at redshift $\sim1$ \citep[e.g.][]{bouche16,zabl19}.

Despite this widespread evidence, we wonder what are the effect of these outflows and inflows on the host galaxy properties.
Several studies have tried to correlate the outflow properties to other galaxy parameters, obtaining contrasting results. 
Some correlations were found between the outflow velocity and SFR-related parameters \citep{heckman15,heckman16,cazzoli16} or the stellar mass \citep{rubin14}. 
In a recent study, \citep{robertsborsani20} conducted an IFU spectral analysis of $\sim400$ massive (\mass $> 10^{10}$ \msun) local star-forming galaxies, finding a correlation between the ISM velocity \vism and \sigmasfr in the central regions. 
\citet{chisholm17} suggest an important effect also from mergers. On the other hand, \citet{steidel10} and \citet{talia12} do not find any correlation with \mass, SFR, or \sigmasfr at redshifts $1.9<z<2.6$ and $\sim 2$, respectively.

It is not yet clear whether the above results depend on an evolution in redshift of these relations or are primarily driven by the mass and SFR ranges probed in each study, with faster and more efficient outflows found when considering extreme starbursting and dusty systems. On the opposite side, as gas outflows are clearly predicted by theory as a consequence of stellar feedback, one hypothesis is that we should go to less massive galaxies (\mass $< 10^{10}$ \msun) to find a significant effect on galaxy properties, given that it is easier for the gas to escape the weaker gravitational attraction in that regime. 
This effect might be stronger as we go to higher redshifts than those analysed statistically in the aforementioned works.
However, it is unclear how the more efficient infall of gas in the early epoch of intense galaxy build up can affect the global ISM kinematics. 

To answer these questions, we exploit the VANDELS survey \citep{pentericci18,mclure18,garilli21}, which in recent years has obtained ultradeep optical spectra for hundreds of galaxies up to redshift $\simeq5$ down to a magnitude of H$_{AB}=27$. The survey has detected the stellar continuum with a high average signal-to-noise ratio (S/N) ($>7$ for the majority of them), and has measured the ISM absorption lines with precision (which are deeper compared to purely photospheric features) even for individual galaxies. Thanks to the long integration times, ranging from $20$ to $80$ hours (depending on the magnitude), VANDELS provides an excellent sample with which to probe the ISM kinematic properties  (using far-UV absorption lines)  of normal star-forming galaxies up to redshift $\sim5$ and down to a stellar mass of $10^{8}$ \msun statistically
 for the first time. 

In \citet{marchi19}, the VANDELS collaboration focused on Ly$\alpha$ emitters from the Data Release 3, finding an anti-correlation between the ISM velocity shift \vism and the Ly$\alpha$ shift, explained as galaxies with higher velocity ISM outflows producing channels for Ly$\alpha$ photons to escape as less affected by scattering processes. This shed light on how Ly$\alpha$ photons are affected by the ISM in star-forming galaxies at redshift $\sim3$. The goal of the present paper is instead to extend and generalise those previous results, focussing on the ISM kinematics of VANDELS galaxies using metal absorption lines in the far-UV rest-frame. We analyse galaxies regardless of their Ly$\alpha$ emission. This yields a better and more representative characterisation of the population of normal star-forming galaxies at $z\sim3$. Moreover, we investigate the presence and properties of outflows and inflows both globally and for individual galaxies, extending the previous studies to lower stellar masses down to \mass of $10^8$ \msun and to higher redshifts up to $z\sim5$. Lastly, we aim to test the correlations of the \vism  with the SFR, mass, and \sigmasfr, and also explore a broader parameter space that includes other physical properties of galaxies, such as morphological parameters. Moreover, following previous findings by \citet{chisholm15}, we also study the gas kinematics in merger and interacting systems at these low masses at $z>2$, comparing the results to more isolated galaxies. 

The paper is organised as follows. In Section~2, we describe the VANDELS spectroscopic survey, the selection of \CIII and \HeII emitters, and the measurement of the ISM velocity shift from far-UV absorption lines. We then present the derivation of stellar masses, SFRs, and other parameters of the galaxies from multi-wavelength broadband photometry. In the last part of the section, we also identify merger systems and measure the galaxy sizes in high-resolution HST images. 
In Section~3, we compare the ISM velocity shifts to multiple physical properties of the host galaxies, both with a spectral stacking approach and on an individual basis. In Section~4, we finally discuss the global physical picture describing gas outflows and inflows in star-forming galaxies at redshift $\sim3$ and compare the results to previous studies on the same topic.  
In our analysis, we adopt a \citet{chabrier03} initial mass function (IMF) and, unless stated  otherwise,  we assume a cosmology with $H_{0}=70$ $\rm km\ s^{-1}Mpc^{-1}$, $\Omega_{\rm m} = 0.3$, $\Omega_\Lambda = 0.7$. We also assume a solar metallicity Z$_\odot$ $=0.0142$ \citep{asplund09}. 

\section{Methodology}\label{methodology}

In this section, we describe the VANDELS spectroscopic survey and the identification of a sample of \CIII and \HeII emitters, for which it is possible to determine the systemic redshift $z_{sys}$ of the galaxies. We then analyse the ISM absorption lines in the far-UV regime, and show the derivation of the ISM velocity shift, with which we can probe the relative motions of the gas, both in inflow and outflow. 
We also present the SED-fitting procedure used to derive the main physical properties, such as stellar mass M$_\star$ and SFR, from the broadband photometry. Finally, we identify merger systems and measure the physical sizes of our galaxies, from which we calculate the SFR surface density.  

\subsection{The VANDELS spectroscopic survey}\label{VANDELS}

We consider spectroscopic data coming from the VANDELS survey, which observed $2087$ galaxies at redshifts $1 < z < 6.5$ with the VIMOS spectrograph at VLT in a period of time between 2015 and 2018. The survey covers two different fields in the sky, namely the Ultra Deep Survey (UDS) and the Chandra Deep Field South (CDFS) fields, totalling an area of $\sim 0.2$ deg$^2$. 
We refer to \citet{mclure18}, \citet{pentericci18}, and \citet{garilli21} for the technical details regarding the design of VANDELS, the preparation of the observations, data reduction, and spectroscopic redshift measurements, while we summarise the main features here. 

The spectra obtained with VIMOS cover the optical wavelength range from $\sim 4900$ \AA\ to $\sim 9800$ \AA, and have a resolution R$\simeq 600$, corresponding to a FWHM$_{res} \simeq 2.7$ \AA\ at $1600$ \AA\ rest-frame. Thanks to the long integration times, ranging from $20$ to $80$ hours per object (depending on its i-band magnitude), most of the spectra have a well-detected stellar continuum, with a S/N per resolution element of higher than $7$ for at least $80\%$ of the sample \citep{garilli21}. The VANDELS team measured the spectroscopic redshift for all the targeted sources with a semi-automatic procedure: the EZ software package \citep{garilli10} was used for a first estimation, while in a second step an independent visual check was made by different team members to confirm the redshift and assign a spectroscopic quality flag in order to keep track of its reliability \citep[for more details see][]{pentericci18}. Spectra with flags $3$, $4,$ and $9$ have the most robust redshift determinations, with $> 95\%$ probability of being correct, as based on the clear detection of emission and/or absorption lines.

\subsection{Line measurements}\label{line_fitting}

In order to detect ISM motions relative to the bulk of the stellar population inside a galaxy, we need an accurate estimation of both the systemic redshift and the ISM absorption line centroids. This is done in our work by fitting the observed far-UV absorption line profiles with Gaussian functions using the Python version of the MPFIT routine \citep{markwardt09}. 
This tool allows the user to estimate all the Guassian parameters of the lines, including central wavelength $\lambda_\text{cen}$, total flux $f_\text{line}$, and RMS width $\sigma_{line}$ (in \AA), with their corresponding uncertainties, while the goodness of the fit is kept under control through the reduced $\chi^2$ ($\chi^2_{red}$) of the fit. 

In all cases, the continuum is parametrised as a straight line and is fitted together with the emission or absorption lines, considering blueward and redward wavelength windows of $\pm80$ \AA. We also impose a lower and upper limit on the line width $\sigma_{line}$ in order to avoid unphysical results where a very broad or narrow line is simply fitting the noise. For the emission lines, we set them to $2$ and $10$ \AA\, respectively, corresponding approximately to $80$ and $400$ km/s depending on redshift. We increase the $\sigma_{line}$ upper limit to $14$ \AA\ when fitting the absorption lines, because we dig down to lower S/N for these lines. In all cases, we do not rely on results where one of the two extreme values of $\sigma_{line}$ is fitted, for which MPFIT also yields a zero uncertainty on the estimated parameter.

For the detected lines, we also estimate their equivalent width (EW) assuming the same linear shape of the underlying continuum. We convert the line widths in velocity space as $\sigma_\text{vel}$ [km/s] $= \sigma_{line}$ / $\lambda_\text{cen}$ $\times\ c$, with $c$ the velocity of light. Alternatively, we also use the observed FWHM throughout the paper, calculated as FWHM [km/s] $=\sigma_\text{vel} \times 2.355$.

\subsection{Systemic redshift estimations for CIII] and HeII line emitters}\label{systemic_redshift_estimation}

\begin{figure}[t!]
    \centering
    \includegraphics[angle=0,width=0.99\linewidth,trim={0.3cm 0.3cm 0.4cm 0.1cm},clip]{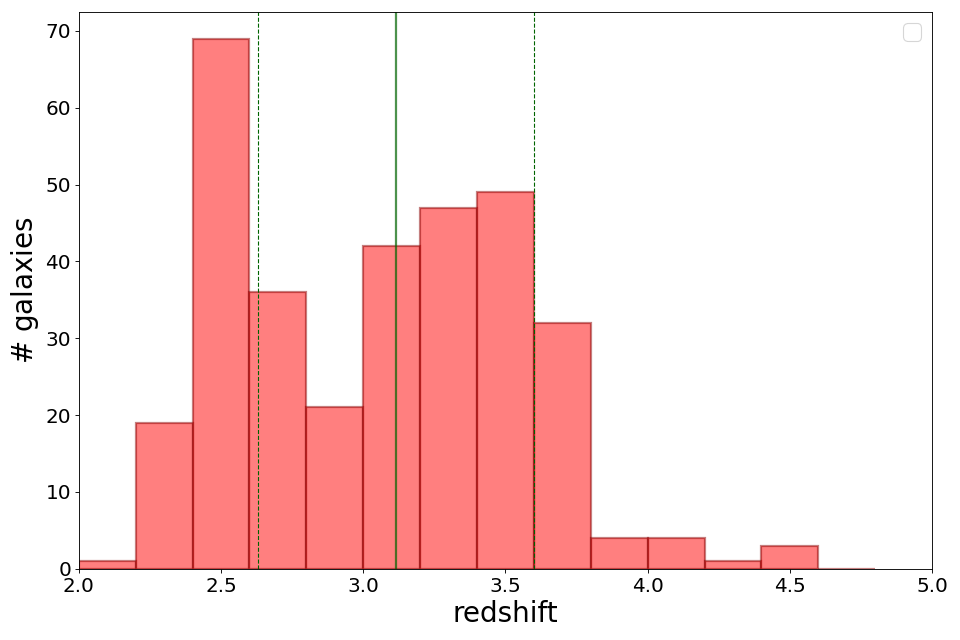}
    \caption{\small Systemic redshift distribution of galaxies with CIII] or HeII emission at S/N $\geq3$, selected in this work as described in the text. The green vertical lines represent the median redshift of the sample (continuous line) and the standard deviation of the distribution (dotted lines).  
    }\label{redshift_distribution}
\end{figure}

Given the wavelength coverage of VANDELS spectra, we use the \CIII line as a tracer of the systemic redshift, which is available up to a redshift of $\sim 4$. 
As the \CIII is a doublet, with vacuum wavelengths of $\lambda\lambda$ $1906.68$ and $1908.73$  \AA, we try first to fit the emission line with a double Guassian, fixing the distance between the two peaks and allowing the ratio to vary between $1$ and $1.6$ \citep{osterbrock06}. 
In practice, given that the lines are unresolved at our resolution, a single Gaussian assuming a central rest-frame wavelength \CIII = ($\lambda_1$+ $\lambda_2$)/2 $=1907.705$ \AA\ already provides a good fit with less parameters for the majority of galaxies.

Indeed, a double Gaussian fit returns a meaningful line ratio only for some cases where the S/N of the CIII] is high (typically $\gtrsim 7$), while the remaining times the code prefers a ratio that is outside of the allowed range, hence returning one of the two extreme values and null uncertainties on the fitted parameters. Comparing the two procedures for a subset with a good double-Gaussian estimation, and running a set of Monte Carlo simulations (described in Appendix \ref{appendix0}), we noticed that the single-Gaussian fit tends to give centroids that are slightly shifted compared to the other method by $\sim-30$ km/s in velocity space, regardless of the S/N of the line if above $3$ (see \ref{appendix0}). 
Therefore, we decided to apply a systematic correction of $+30$ km/s when deriving the systemic redshift through a single-Gaussian fit of \CIII (i.e. a correction of $-30$ km/s to the velocity shifts when comparing to the systemic frame). This might be due to the first, brighter component of the \CIII doublet, which weighs more when fitting the emission profile with a single Gaussian, slightly shifting the centroid to the blue side. 

When the \CIII is not detected or when it does not fall inside the VIMOS wavelength coverage, we look for the \xHeII emission with S/N $\geq 3$, which can be considered as an alternative tracer of the systemic redshift \citep{saxena20}. The \HeII emission is always fitted with a single Gaussian.
When both \CIII and \HeII lines are detected with S/N $\geq 3$, we consider the redshift inferred from \CIII, as the \HeII line has typically a lower S/N and a larger uncertainty in the estimated Gaussian parameters. 
Comparing the ISM-shift values derived assuming the \CIII and the \HeII lines as systemic redshift indicators, respectively, we find no evident systematic offsets for galaxies where both lines are reliably detected with S/N $\geq 3$ (see Appendix \ref{appendix1}), which supports our choice of using \HeII to infer $z_{sys}$ as an alternative to \CIII. We also note that, owing to the maximum FWHM allowed for the \HeII in the fit ($\simeq1000$ km/s), we do not expect a significant contribution from Wolf-Rayet stars \citep{shirazi12}. 
We finally remark that the stellar photospheric absorption lines, which would also probe the systemic redshift, are too faint to be detected in individual objects.

Among our sample of star-forming galaxies at redshifts $\geq2$, we detect the \CIII line with a S/N of larger than $3$, and estimate \zsys from this line for $276$ galaxies; for $6$ of these we use the double component Gaussian fit. For $73$ galaxies, \zsys is determined from the HeII line only, among which $9$ are at redshift $>3.9$, and the remaining $64$ are at lower redshift but with undetected \CIII. 
This procedure yields $349$ \CIII or \HeII line emitters in total, for which it was possible to estimate their systemic redshift. 

\subsection{Identification and exclusion of AGNs}\label{AGN_exclusion}

\begin{figure*}[ht!]
    \centering
    \includegraphics[angle=0,width=1\linewidth,trim={0.3cm 0cm 0.2cm 0.1cm},clip]{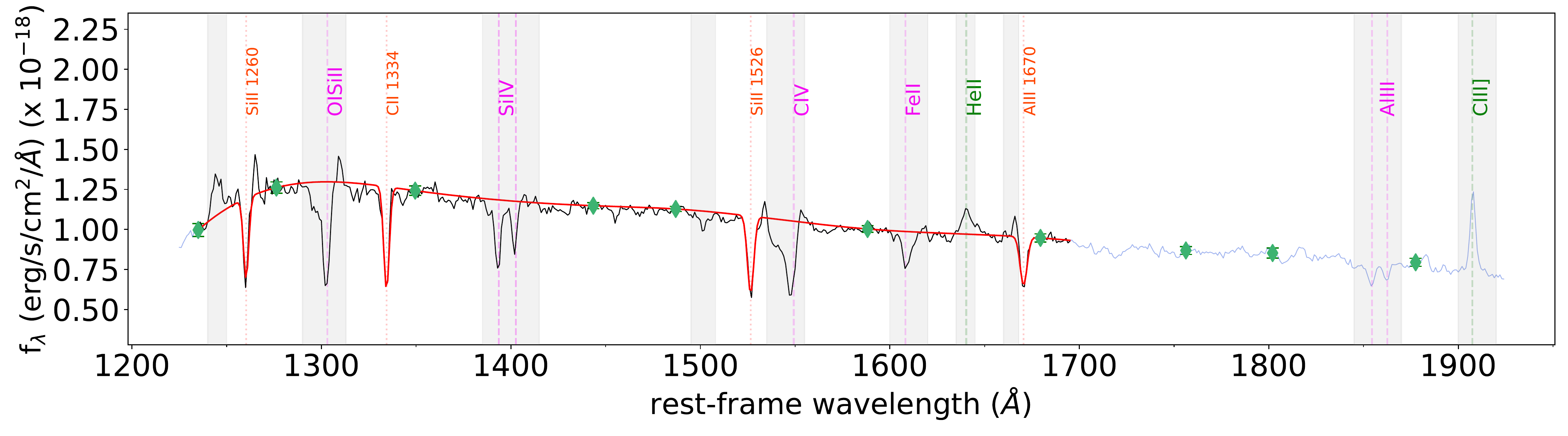}
    \caption{\small Stack of all star-forming galaxies selected for our analysis (AGNs excluded). The green diamonds represent the pseudocontinuum points used for fitting the cubic spline to the stellar continuum. The red continuous line highlights the combined fit to the \xSiIIa, \xSiIIb, \xCII, and \xAlII absorption lines, while the grey shaded regions have been masked in the fit. The vertical lines highlight the features used in the combined fit (in red), the absorption lines excluded from the combined fit but measured separately (in pink), and the emission lines used for the systemic redshift estimation (in green).
    }\label{figure_stack_starforming}
\end{figure*}

As our goal is to study the presence and effects of outflow activity induced by star formation, we have to identify and remove the contribution from AGNs.  
AGN candidates are selected based on multiple criteria. Briefly, AGNs were identified according to their X-ray emission, radio emission, or UV-based emission line diagnostic diagrams. 

In particular, X-ray AGNs are identified by cross-matching the position of our VANDELS sources with the Chandra-based X-ray catalogues of \citet{luo17} and \citet{kocevski18} (in CDFS and UDS, respectively), imposing a matching radius of $1.5''$. The AGN catalogues are complete above an X-ray luminosity of $\sim 10^{42.5}$ ($\sim 10^{43}$) $erg/s$ in CDFS (UDS) up to a redshift of $\sim 4$, which is the limit for most of the galaxies considered in this work.  We also visually check the centroid of the X-ray emission from Chandra superimposed on the high-resolution HST-ACS i-band cutouts to exclude wrong identifications due to nearby optical systems. 
Radio AGNs are identified using the radio source catalogues by \citet{simpson06} and \citet{miller13}. 

Finally, UV AGNs are identified by first looking at the spectra for strong \CIV emission, with a similar approach to that adopted by \citet{saxena20}. For these \CIV emitters, we then identify potential AGNs by comparing the \CIV/\HeII ratio to \CIV/\CIII, and exclude those sources that lie in the AGN region of the diagram according to the photoionisation models of \citet{feltre16} (see their Fig. 5).  Considering the average S/N of our spectra, these are the only bright emission lines that could be detected in individual galaxies and that are used to separate AGN-driven and star-formation-driven radiation. 

From this procedure, we identify and exclude $19$ AGN candidates, of which $17$ have AGN-like emission line ratios, $8$ are detected in X-ray, and $1$ in radio. An alternative diagnostic diagram comparing the EW(\CIV) to \CIV/\HeII ratio ---which is based on the modelling of \citet{nakajima18} and can distinguish between star-formation- and AGN-driven ionisation--- yields the same sample of AGN candidates. 
We note that more details on the UV-based selection and a discussion of the full VANDELS AGN sample, including those at redshifts lower than $2$ and X-ray- or radio-selected AGNs that do not emit \CIII or \HeII, will be presented in a forthcoming paper of our collaboration (Bongiorno et al., in preparation). 

We are thus left with a final sample of $330$ purely star-forming galaxies with a reliable systemic redshift estimation from \CIII or \HeII. For this subset, $34$ galaxies have $z_{sys}$ estimated from \HeII. The redshift distribution of the final sample is shown in Fig. \ref{redshift_distribution} and ranges from $2$ to $4.6$, with the bulk of the population comprised between $2.2$ and $3.8$.

\subsection{Stacking analysis}\label{stacks}

Even though we can detect and measure ISM velocity shifts for individual galaxies, we also perform spectral stacking to better test the correlations among the different physical parameters. 
The advantages of stacking are manifold. Firstly, it provides increased statistics as we also include in the stack objects where some ISM absorption lines are undetected. Secondly, we significantly increase the S/N of the stellar continuum, allowing to us check for the presence of faint, broad, or asymmetric wings in the ISM absorption line profiles, which could be indicative of additional outflows or inflows. Furthermore, in the stacking, we can study all the absorption lines in the rest-frame spectral range probed by VANDELS  simultaneously, while this is not possible for every individual galaxy, depending on the redshift of each. Finally, we also reduce the uncertainty associated to the ISM shift measurement, and visualise how it is related to the other galaxy properties  globally. 

The first step of the stacking procedure is the conversion of all our spectra to the rest frame using the systemic redshift estimated in Section \ref{systemic_redshift_estimation} from the \CIII or \HeII emission lines. The spectra are then normalised to the median flux in the range $1570$-$1601$ \AA\ and resampled to a wavelength grid of $1$ \AA\ per pixel following the method of our previous works \citep[e.g.][]{calabro22}. 
Finally, the composite spectra are derived taking the median flux in each pixel wavelength after applying a $3\sigma$ clipping to remove outliers. The uncertainty on the stacked spectrum was instead calculated with a bootstrapping resampling procedure as in \citet{calabro22}. 

We also tested another procedure by taking the weighted average flux in each wavelength pixel to derive the composite spectrum, where the weight is given by the S/N of the emission line (either \CIII or \HeII) that was used for the systemic redshift estimation. This way, spectra with a more accurate estimation of \zsys contribute more to the final composite. However, we find that this approach does not lead to significantly different results and therefore we adopt in this paper the first procedure. This should avoid the introduction of subtle systematic biases due to the different contribution of each type of galaxy to the final spectrum, which would be difficult to quantify and control. A representative, high-S/N ($\simeq100$) stacked spectrum of all star-forming \CIII and \HeII emitters selected in the previous section is presented in Fig. \ref{figure_stack_starforming}.

\subsection{Probing the ISM kinematics with absorption lines}\label{ISMabsorption}

\begin{figure*}[ht!]
    \centering
    \includegraphics[angle=0,width=1\linewidth,trim={0.cm 0.cm 0.cm 0.cm},clip]{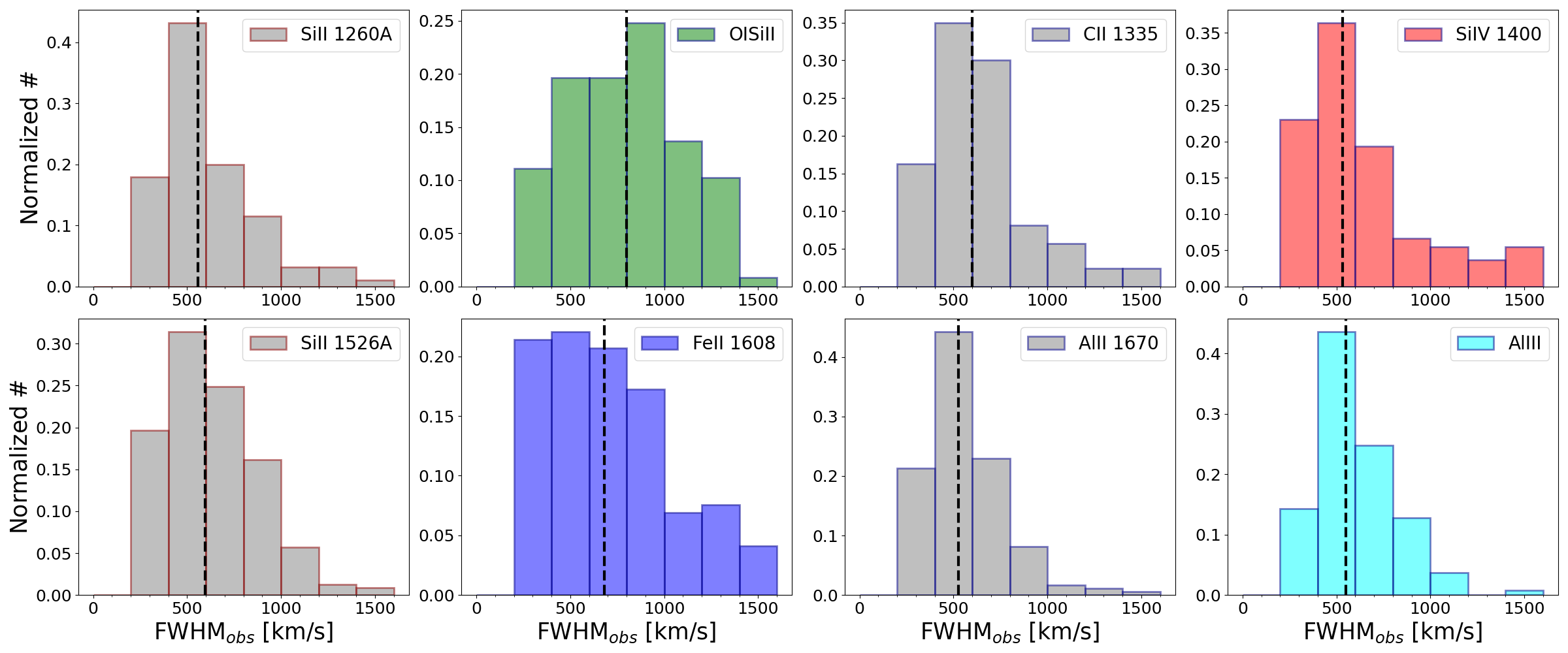} 
    \caption{\small Distribution of FWHM (in km/s) for the sample selected in this work, for the following absorption lines detected with a S/N $\geq2$: \SiIIa, \OISiII, \SiIIb, \CII, \FeII, \AlII, \SiIV, and \AlIII. The vertical lines highlight the median FWHM for each of the lines analysed.}
    \label{histograms_sigma}
\end{figure*}

\begin{figure*}[ht!]
    \centering
    \includegraphics[angle=0,width=1\linewidth,trim={0.cm 0.cm 0.cm 0.cm},clip]{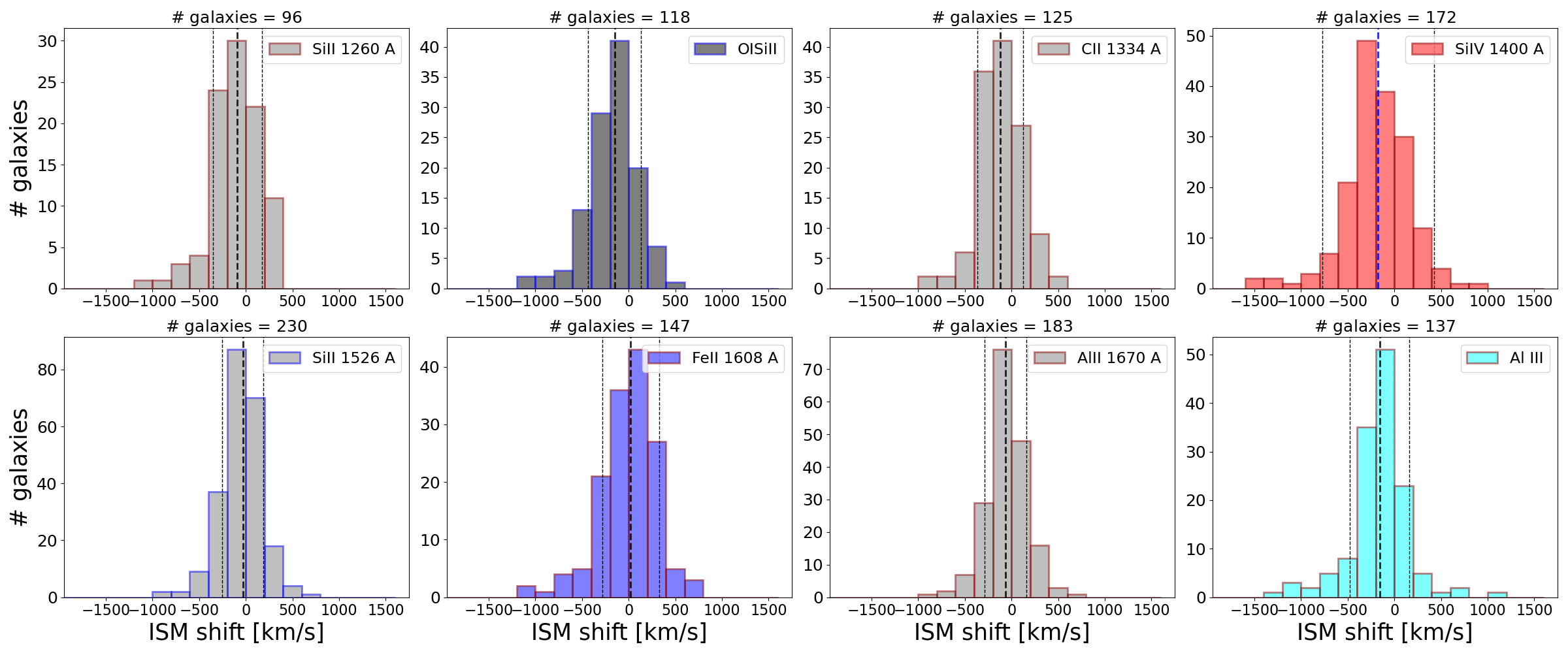} 
    \caption{\small Distribution of ISM shift for the same lines shown in Fig. \ref{histograms_sigma}. Symbols and colours are the same as in the previous plot. The two dotted lines in addition to the dashed line of the median represent the standard deviation of \vism values. The number of galaxies contributing to each histogram is written in the top of each panel.}
    \label{histograms_ISMshift}
\end{figure*}

\begin{figure}[ht!]
    \centering
    \includegraphics[angle=0,width=1\linewidth,trim={0.cm 0.cm 0.cm 0.cm},clip]{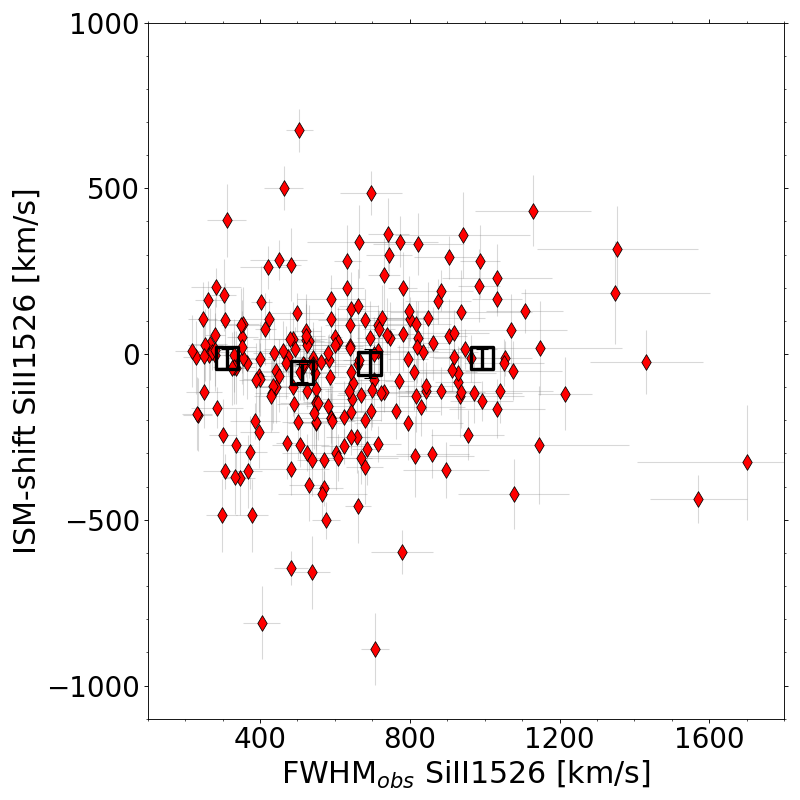} 
    \caption{\small Scatter plot comparing the ISM velocity shift of \xSiIIb to the FWHM of the same line (red diamonds). The black squares are the median \vismSiIIb calculated in four bins of FWHM. In both cases, it is clear that there is no correlation between the two quantities. A similar result is also obtained for all the other absorption lines.}
    \label{scatter_sigma_ISMshift}
\end{figure}

\begin{table}[h!]
\centering{\textcolor{blue}{ISM absorption lines}}
\renewcommand{\arraystretch}{1.5}  
\vspace{-0.2cm}
\begin{center} { \normalsize
\begin{tabular}{ | m{2.8cm} | m{4.7cm} | } 
  \hline
  \textbf{Absorption line} & \textbf{central wavelength} \\ & \textbf{(rest-frame vacuum) [$\AA$]} \\ 
  \hline
  \textcolor{red}{\SiIIa} & \textcolor{red}{$\lambda\; 1260.422$} \\  
  \OISiII & $\lambda\lambda\; 1302.168$-$1304.370$ \\
  \textcolor{red}{\CII} & \textcolor{red}{$\lambda\; 1334.532$} \\
  \SiIV & $\lambda\lambda\; 1393.755$-$1402.770$ \\
  \textcolor{red}{\SiIIb} & \textcolor{red}{$\lambda\; 1526.707$} \\
  \CIV & $\lambda\lambda\; 1548.202$-$1550.774$ \\  
  \FeII & $\lambda\; 1608.45083$ \\
  \textcolor{red}{\AlII} & \textcolor{red}{$\lambda\; 1670.7874$} \\
  \AlIII & $\lambda\lambda\; 1854.716$-$1862.79$ \\
  \hline
\end{tabular} }
\end{center}

\caption{\small Table indicating the absorption lines that we detected and fitted in this work, with their rest-frame wavelengths used for the derivation of the ISM velocity shifts. The low-ionisation absorption features that are fitted simultaneously in the combined fit are shown in red.}
\label{tabella1}
\end{table}

After calculating the systemic redshift, we proceed to study the information on the ISM kinematics that can be inferred from the far-UV absorption lines. 
Here we have more options, that is, a larger number of transitions that we can use to assess the ISM properties. 
The ISM absorption lines that we detect in individual spectra in the wavelength range from $1000$ to $2000$ \AA\ are presented in Table \ref{tabella1} and can also be visualised in Fig. \ref{figure_stack_starforming}. 

We note the presence of lower and higher ionisation lines. In the first category reside the \xSiIIa and \xSiIIb \AA\ absorption lines (dubbed \SiIIa and \SiIIb in the rest of the paper), \xCII, \xAlII, \xFeII \AA\;, 
and \xOISiII \AA\ (rest-frame vacuum wavelengths). These low-ionisation lines (LIS) mostly trace the neutral and low-ionisation gas in and around galaxies \citep{shapley03}. 
With  the exception of \OISiII, all the lines are fitted with the same procedure adopted for the emission lines, that is, with a single Gaussian in absorption plus an underlying continuum modelled with a straight line using windows of $80$ \AA\ blueward and redward of the lines. 

Given that these windows may overlap with other bright emission or absorption features, when estimating the continuum, we mask the spectral regions corresponding to bright emission lines such as Ly$\alpha$, \NV, \xCIV, \xOIII, \xCIII, and \xHeII, and the broad absorptions due to \xOISiII (when not fitting this line) and \xCIV. 
This is more important for the spectral stacks and for the \xSiIIa and \xAlII lines. 
The \OISiII feature is instead modelled as a double Gaussian, fixing the wavelength separation among them. An alternative fitting with a single Gaussian at the median wavelength provides a poorer fit in general, although the results are not significantly affected. 

In addition to the LIS, there are also absorption features from higher ionisation species, namely the doublets \xSiIV and \xAlIII, where the latter is in general fainter and detected for a lower number of systems, and will therefore be studied systematically only in the composite spectra. In spite of their ISM origin, they also have a stellar wind component which usually manifests as a P-Cygni profile due to the contribution of very young massive stars. 
The two doublets were fitted with a double Gaussian, fixing the relative wavelength separation and imposing the same velocity width for the two components, but leaving the ratio free to vary between $1$ and $1.6$, which is the physically allowed range given the electronic densities of local star-forming regions \citep[e.g.][]{osterbrock06}. 

In Fig. \ref{histograms_sigma} we show the distribution of the FWHM for the absorption lines introduced above. 
In general, \xSiIIa and \xFeII are the lines detected more frequently in our sample. Besides the fact that they are among the deepest absorption features, the main reason for this is that they lie in a wavelength range covered by VIMOS over almost all of our redshift range, from $z=2$ to $5$. 

Overall, the individual observed FWHM values range between $200$ and $1800$ km/s, while the medians are all within the range of $550$-$700$ km/s. The distributions are rather similar, with a $1\sigma$ dispersion of $200$-$300$ km/s around the median values, depending on the line. In particular, \FeII is the broadest line, with a median FWHM of $680$ km/s. 
This might be due to a significant absorption component of \FeII at the systemic redshift, and to a contribution from secondary transitions redward of the main \FeII ion line at $1608.45$ \AA. 

At this point we can calculate, for all the ISM lines seen above, the ISM velocity shift (i.e. ISM-shift, or simply \vism), defined as:
\begin{equation}
v_{IS,\ line}= \frac{z_{line}-z_{sys}}{(1+z_{sys})} \times c
,\end{equation}

where $c$ is the velocity of light, $z_{line}$ is the redshift of each line derived from the best-fit Gaussian centroid, and $z_{sys}$ is the systemic redshift. The uncertainty is derived from the error propagation formula. 
In Fig. \ref{histograms_ISMshift} we show the distribution of \vism for each of the absorption lines introduced before. In general, the ISM-shift ranges between $-1000$ km/s and $500$ km/s, meaning that we can have signatures of both gas outflows (negative \vism) and inflows (positive \vism), with a median value that is typically within $-200$ and $+50$ km/s, depending on the line considered.

We checked how the \vism and the FWHM of the lines are related, and we find no correlation between these two quantities. Figure \ref{scatter_sigma_ISMshift} shows an example for the \xSiIIb line, which is available for more galaxies in our sample, even though a similar result is also found for the other lines, and for the \HeII and \CIII in emission.

In Appendix \ref{appendix2}, we compare \vism inferred from multiple absorption lines (see Fig. \ref{fig_app_1} and \ref{fig_app_2}). In particular, we find that the \SiIIa, \SiIIb, \CII, and \AlII lines have similar properties in terms of their FWHM and ISM-shift distributions, which are well aligned along the 1:1 correlation with no evident systematic offsets (Fig. \ref{fig_app_2}).
On the other hand, the \FeII line, in addition to having a broader velocity width distribution, shows a positive velocity offset by $100$-$150$ km/s compared to the other lines mentioned above, and its median is more consistent with the systemic redshift. However, we caution against the use of the \FeII as a systemic redshift indicator for the parent galaxy population because of the large dispersion in \vismFeII for individual galaxies, which also makes the derivation of a systematic correction somewhat uncertain. 
Although we find a median offset of $+40$ km/s in our sample, we find a significant number of systems with \FeII in outflow or inflow up to $\pm1000$ km/s from the central value. 

The higher ionisation lines, that is, \AlIII and \SiIV, tend to have lower median velocity shifts compared to the low ionisation lines (by $50$ and $110$ km/s, respectively), suggesting larger outflow velocities of the high-ionisation gas, which we discuss further below. 
Remarkably, we also see that the \AlIII and \SiIV absorption shifts (\vismAlIII and \vismSiIV) are instead tightly correlated, with no systematic offsets with respect to the 1:1 line.
Finally, the \OISiII absorption feature also tends to give higher outflow velocities on average compared to the \xSiIIb line by $\sim 100$ km/s. In this case, the difference might be related to the complex shape and doublet nature of the absorption where the contribution of the two elements in different ionisation states may vary from case to case. 

\subsection{The combined fit of low-ionisation lines}\label{combined}

The results shown in the previous section indicate that the \SiIIa, \SiIIb, \CII, and \AlII share similar properties, and therefore in principle they can be fitted together in order to give a unique, combined, and more precise estimate of the low-ionisation and neutral ISM velocity shift. 
In the combined measurement, we first model the continuum by fitting a cubic spline to the pseudocontinuum ranges identified by \citet{rix04} as these relatively free from strong absorption and emission lines, as already done in \citet{calabro21}. To these, we add another window blueward of the \xSiIIa feature in the rest-frame range $1233.0$-$1237.0$ \AA\  in order to better fit the bluest absorption feature in our spectrum. We then simultaneously fit a single Gaussian to all the individual absorption lines (among the four listed above) that have a S/N of at least $2$, fixing the relative distance among them and their velocity width $\sigma$. 
An example of this procedure is shown in Fig. \ref{figure_stack_starforming} for the whole sample of star-forming galaxies with a good estimate of \zsys. We derived \vismcomb  using all four
lines in $50$  galaxies, using three lines in 127 galaxies, using two lines in 67 galaxies, and using one low-ionisation
line with S/N $\geq$ 2 in  57 galaxies.  The output FWHM and ISM velocity shift from the combined fit agree overall with the typical values found when using the individual features separately and when using the average or the median of the lines available. 

To further check which lines should be considered in the combined fit, we also run a combined measurement including all possible far-UV absorption lines in the fit (both those in pink and in red in Fig. \ref {figure_stack_starforming}), and then calculating the number of outliers for which the combined ISM velocity shift differs from the individual line estimation by more than $200$ km/s. 
The result of this exercise is shown in Fig. \ref{outliers_test} in Appendix \ref{appendix2}. Imposing an outlier threshold of $10\%$, \OISiII, \SiIV, \FeII, and \AlIII are automatically excluded, while the remaining lines (\SiIIa, \SiIIb, \CII, and \AlII) have outlier fractions of less than $5\%$, confirming their similar nature. 
Throughout this paper we therefore use and refer to the combined fit as a single, simultaneous fit to all the available low-ionisation lines among the four listed above. 

\begin{figure*}[t!]
    \centering
    \includegraphics[angle=0,width=0.95\linewidth,trim={0.cm 6.cm 20.cm 0.cm},clip]{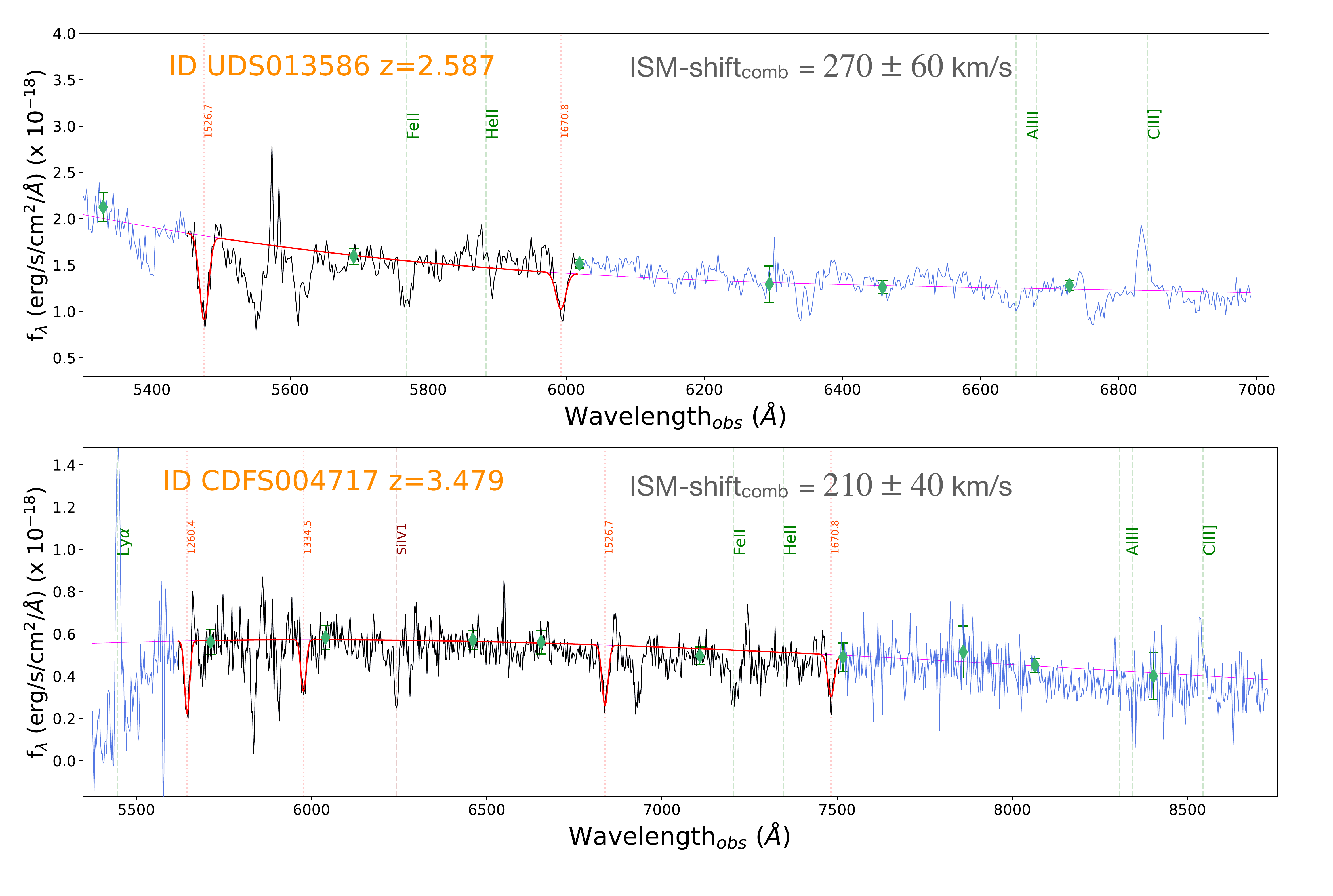} 
    \vspace{-0.5cm}
    \caption{\small Figure shows the spectrum of two galaxies at redshifts $2.587$ and $3.479$ in the UDS and CDFS field (top and bottom panel, respectively) with the combined fit of low-ionisation absorption lines (red), and for which we detect significant inflow signatures, with positive ISM velocity shifts as indicated in each panel (i.e. the centroids of ISM absorption lines are redshifted compared to the \CIII centroid used for the systemic redshift). 
    }\label{Figures_inflow}
\end{figure*}

While for the spectral stacking we use the whole sample of $330$ star-forming galaxies selected as \CIII or \HeII emitters with a good estimation of the systemic redshift, the ISM absorption lines (and therefore a measurement of the ISM velocity shift) are available for a smaller number of objects than the original one. 
In particular, we have a combined fit measurement of the low-ionisation absorption lines (\vismcomb) for $299$ galaxies, while the \SiIV line (and therefore an estimate of \vismSiIV) is available for $166$ objects. We use these two measurements in the results section to study the low-ionisation and high-ionisation gas kinematics, respectively, as a function of other physical properties of our sample.
The exact number of galaxies for which a particular ISM absorption line is detected with S/N $\geq 2$ is instead indicated above each histogram in Fig. \ref{histograms_ISMshift}. We also note that choosing a higher S/N threshold for the inclusion of a low-ionisation absorption line in the combined fit does not change the results. Moreover, we should keep in mind that the combination of multiple features fitted at the same redshift significantly reduces the possibility of spurious detections, meaning that we can be confident of the presence of the absorption lines even if we decrease the S/N requirement on the single feature to $2$.

\subsection{The maximum ISM velocity}\label{vmax}

While the bulk velocity describes the global kinematic properties of the ISM, in reality the gas component in a galaxy shows a range of velocities, which can translate into narrower or broader absorption line widths. In order to understand the final fate of the gas, it is also useful to constrain the maximum velocity at which the gas is flowing outwards, indicated as \vmax. 

This quantity can be derived following the methodology of \citet{zakamska14}. In the most general, non-parametric approach, it is based on the cumulative velocity distribution as $F(v)=\int_{-\infty}^vf(v'){\rm d}v'$, where $f(v)$ is the best-fit spectrum modelled around the absorption feature and translated into velocity space.  
We then define \vmax as the velocity (always reported to the systemic redshift) at which $2\%$ of the total flux of the line accumulates, which analytically is the solution to the equation $F(v)=0.02$, if $F(v)$ is normalised to the total flux. This also corresponds to $v_{02}$ which is typically used in the literature. 

In this work, the lines are well described as single Gaussians, in which case \vmax is simply related to the line FWHM and can be calculated as
\begin{equation}\label{Eq_vmax}
v_{max} [km/s] = 0.872 \times FWHM_{line} + v_{ISM,line} ,
\end{equation}
where $v_{ISM,line}$ is the velocity shift of the line centroid itself with respect to \zsys, and FWHM$_{line}$ is the intrinsic line width deconvolved from the instrumental broadening.
Furthermore, given the spectral resolution of VANDELS, the absorption lines are marginally resolved, and therefore we need a high S/N to accurately
measure the intrinsic FWHM. For this reason, we only apply this analysis to the spectral stacks in Section \ref{results2}. 

\subsection{Galaxies with positive ISM velocity shift}\label{galaxies_with_inflows}

An important piece of evidence from the histograms presented in Fig. \ref{histograms_ISMshift} is that a small but significant number of sources, depending on the specific line considered, shows a positive ISM velocity shift, which is indicative of a global inflow. This result is confirmed with the analysis of the combined fit, suggesting that positive velocities are not due to noise or found only for specific lines in the galaxies, but rather that all the low-ionisation lines have consistent kinematics. 
In particular, we find that $34 \%$ of our sample has \vismcomb $\geq 0$. We analyse the physical properties of this subset (e.g. \mass, SFR, and morphology) below, and also compare to the global population.

To check the robustness of this result, we selected only galaxies for which the low-ionisation absorption lines that contribute to the combined fit have a S/N $>3$ instead of $2$. This way, we find that the fraction of objects that still have \vismcomb $\geq 0$  is $32 \%$. Even by increasing the detection threshold of the \CIII line used for systemic redshift estimation  to 5, we still get a fraction of $35\%$. 
For explicative purposes, Fig. \ref{Figures_inflow} shows  two examples at redshifts $\sim 2.5$ and $3.5$ of galaxies where we detect positive low-ionisation ISM velocities of significantly above zero.

\subsection{Fitting the SiIV doublet}\label{extra_outflows}

\begin{figure}[t!]
    \centering
    \includegraphics[angle=0,width=0.86\linewidth,trim={0.cm 2.cm 5.cm 0.cm},clip]{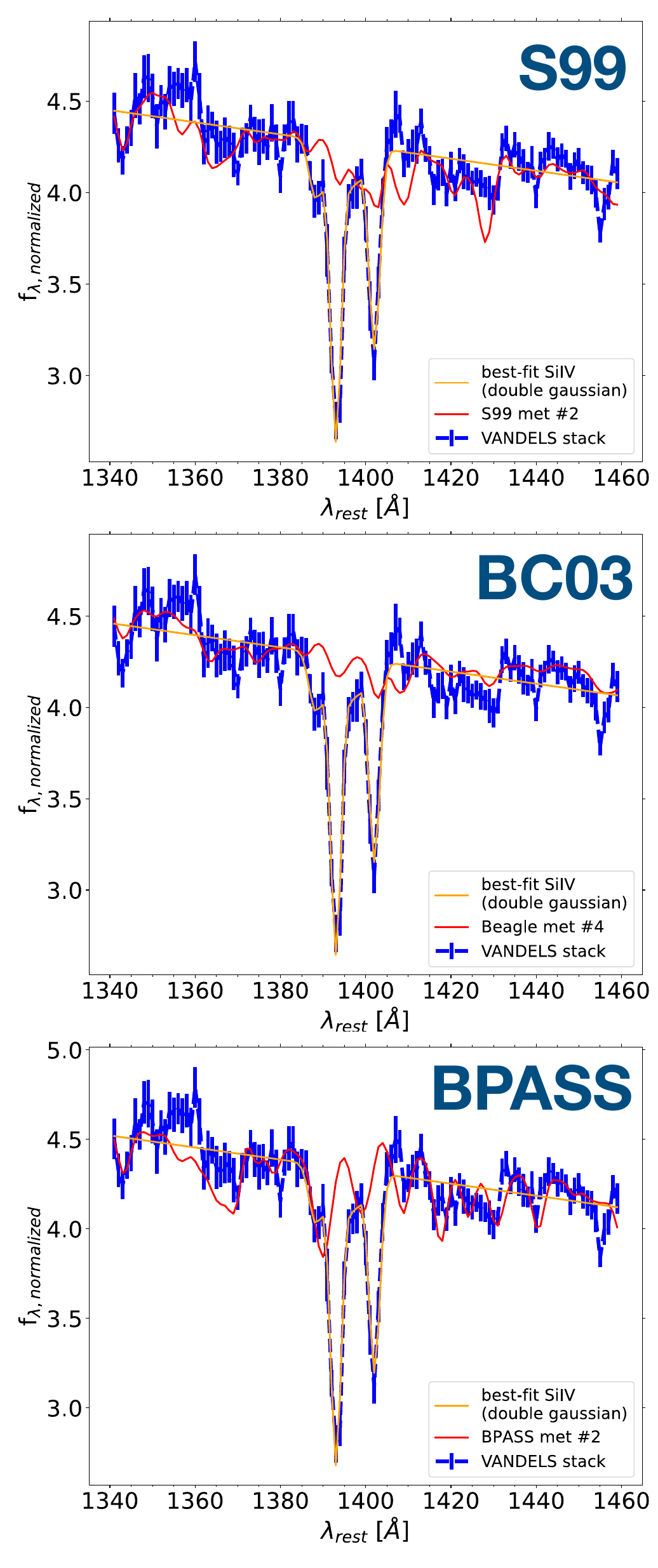} 
    \vspace{-0.3cm} 
    \caption{\small Spectral region around the ISM+stellar \SiIV absorption doublet. In all the panels, the blue line is the stacked spectrum of all star-forming galaxies with \CIII or \HeII emission selected in this work, where the error bars at each pixel represent the $1\sigma$ error. The orange line is the best-fit Gaussian made with MPFIT, assuming two Gaussian components for each side of the doublet (i.e. four Gaussians in total), as explained in the text. The red lines represent the best-fit models from Starburst99 (S99), Beagle (i.e. BC03 models), and BPASS, from top to bottom, respectively. The best-fit models have the closest stellar metallicity Z$_\star$ to the median of our galaxies \citep{calabro21}, that is, $\log$ (Z$_\star$/Z$_\odot$) $=-0.7$ for S99, $-0.82$ for BC03, and $-0.85$ for BPASS. We reiterate the fact that the measured Z$_\star$ using BPASS calibrations is $0.1$ dex lower than S99.}\label{SiIV_fitting}
\end{figure}

A single component Gaussian yields a good fit to the absorption line profiles for most of the galaxies in our sample. However, we notice that in some cases, the \SiIV feature, which has among the largest absorption equivalent widths and is therefore easier to detect even in galaxies with a modest S/N continuum, shows a residual absorption in the bluer part. This suggests that an additional blueshifted component should be added in the fit.    

In order to identify these cases, we systematically fitted all the spectra in the \SiIV range with a double component, which ---as this line is already a doublet--- translates into four Gaussians in absorption fitted simultaneously. In this fit, we imposed that the two \SiIV Gaussians of the extra component have the same velocity width, as in the two main \SiIV absorptions. Moreover, we set a maximum FWHM for the new component to three times the FWHM of the main one, and a maximum velocity difference between the two of $2000$ km/s in order to avoid unrealistically broad absorption features. 
Finally, we compared the reduced $\chi^2$ ($\chi_{red}^2$) values obtained through a single component fit with those given by the new procedure. We then selected the galaxies where the additional outflow component flux has a S/N of at least $3$, and, following \citet{zakamska14}, where the double component fit decreases the $\chi_{red}^2$ by $\geq 5\%$ compared to the previous approach. We obtain a total of 22 galaxies satisfying all the above requirements and showing evidence of asymmetric profiles of \SiIV. The main Gaussian component, which might be already shifted with respect to the systemic redshift, is closer to the lower ionisation line velocity shift, while the second is more blueshifted by on average $1000$ km/s. We also notice that such an additional Gaussian component is always fainter than the main \SiIV absorption, with a typical flux ratio of $0.05$. 

The \SiIV doublet has in general a complex shape because, in addition to the ISM absorption, it can include  P-cygni-like, metallicity-dependent, stellar wind features produced by O-type stars \citep{castor75,kaper92}. This P-cygni profile, with a strongly blueshifted absorption component, becomes stronger at older ages or at higher stellar metallicities \citep{drew89,pauldrach90}. 
For this reason, the entire \SiIV observed profile, including the fainter blueshifted absorption wing, is difficult to interpret. It may be an additional outflow component, representing gaseous clouds with a higher bulk velocity compared to the main, deeper absorption measured from a single Gaussian fit. Alternatively, it can be related to stellar physics.

In order to search for the correct hypothesis, we compare the observed \SiIV absorption feature with multiple stellar models, including Starburst99 \citep[S99,][]{leitherer10}, BPASS with binary stars \citep{eldridge17,stanway18,xiao18}, and Beagle \citep{chevallard16}, which is based on the latest version of \citet{bruzual03} stellar population models. For this exercise, we take the spectral stack of all star-forming galaxies selected in this work in order to have the highest possible S/N around the \SiIV feature and detect even the finest stellar features. 


The spectral comparison is shown in Fig. \ref{SiIV_fitting}, where the best-fit model is chosen using a $\chi^2$ minimisation approach and always corresponds to the closest stellar metallicity Z$_\star$ derived for star-forming galaxies at similar redshifts in previous VANDELS works \citep[see][]{calabro21}. The models also have a stellar age of $\simeq 100$ Myr. We can see that S99 and BC03 models, while nicely reproducing the shape of the stellar continuum on the two sides of the \SiIV feature, are mostly flat on top of the ISM absorption feature and do not reproduce the additional \SiIV component. 
On the other hand, the BPASS model considered in the last panel (with a metallicity $\log$ (Z$_\star$/Z$_\odot$) $=-0.85$) can almost perfectly reproduce the extra blueshifted absorption component of \SiIV, which is displaced by exactly the same amount (namely $\sim 1000$ km/s) as the observed difference between the main \SiIV absorption peak and the best-fit centroid of the extra component. 

Furthermore, we find that the additional blueshifted \SiIV absorption is preferentially found in galaxies with higher stellar masses and SFRs, but is also detected when stacking together galaxies at lower \mass and SFR, or all those that do not show this extra component individually. 
This result indicates that we are dealing with an intrinsic physical property of the \SiIV line arising from stellar winds of more evolved O-type stars, and predicted only by the BPASS models. The reason for this might be related to a combination of the inclusion of binary stars and the different modelling of the stellar evolution. 
The detection of this feature for only a small subset of galaxies might also be related to the different stellar ages of individual systems, where more evolved galaxies might have a stronger \SiIV P-Cygni feature, according to BPASS models. 
However, it can be easily detected when stacking together multiple systems, because we are both increasing the S/N of the continuum and averaging among a wide range of stellar ages.  


In the following part of the paper, we refer to the \SiIV line as the component originating in the interstellar medium only. Therefore, in the spectral stacks that we show in Sections \ref{results1}-\ref{results2}, and for the $22$ galaxies
mentioned above where the extra blueshifted component was significantly detected, we always perform a four-Gaussian fit to decontaminate the absorption profile from the contribution of the stellar wind, and measure peak wavelength, total flux, and FWHM only for the ISM component. Given the low equivalent width of this additional feature and the displacement of $\sim 1000$ km/s of its peak, it does not significantly affect the properties of the main \SiIV absorption when the extra-component is not detected.

We also checked in the above subset for the simultaneous presence of blueward asymmetric profiles in other lower ionisation absorption lines. While for the \AlII feature it is difficult to make this test due to the nearby emission of the \xOIII line and \AlIII has on average a much lower S/N in individual galaxies, we do not find evidence of bluer or redder additional components in the \SiIIa, \SiIIb, or \CII lines for our galaxies. This again suggests that the blue wing of the \SiIV feature does not represent an additional outflow component, which we would otherwise detect also in the other lines, at least in the stacks. 


\subsection{Galaxy physical properties from SED fitting}\label{SEDfitting}

\begin{figure}[t!]
    \centering
    \includegraphics[angle=0,width=1\linewidth,trim={0.3cm 0cm 0.2cm 0.1cm},clip]{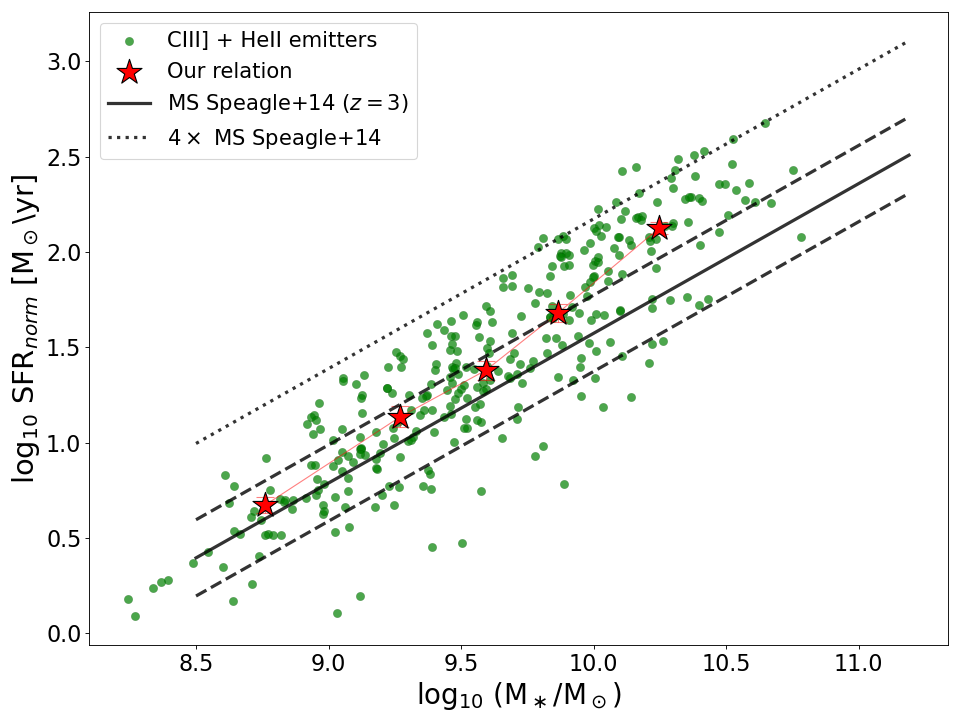}
    \vspace{-0.1cm}
    \caption{\small Star-formation rate--stellar mass diagram for the sample of $330$ \CIII + \HeII emitters selected for our analysis. The main sequence relation by \citet{speagle14} is drawn with a black continuous line, while the dashed parallel lines represent the $1\sigma$ dispersion and the dotted line the starburst limit ($4 \times$ above the main sequence). Our best-fit \mass--SFR relation is represented with large red stars. On the y axis, we consider the SFR normalised to $z=3$ assuming the evolving trend with redshift as $(1+z)^{2.8}$ from \citet{sargent12}.
    }\label{diagram_Mass_SFR}
\end{figure}

For the entire sample of galaxies selected in the previous section, we infer their fundamental physical properties, including stellar mass (M$_\star$) and star-formation rate (SFR), from multi-wavelength photometric catalogues available in the VANDELS fields. In the central parts of CDFS and UDS, which is covered by CANDELS \citep{grogin11,koekemoer11}, the photometry is taken from \citet{guo13} and \citet{galametz13}, respectively, which reach H$_{AB}$ $\sim25.5$ and $\sim26.7$ magnitudes at $\sim 90 \%$ completeness. For the extended region outside CANDELS, our collaboration produced new photometric catalogues from ground-based optical and near-infrared observations, reaching a depth of $H_{AB}=$ $24.5$ and $25$ magnitudes in the CDFS and UDS fields, respectively.

After assembling the whole catalogue, we use the Beagle software (BayEsian Analysis of GaLaxy sEds) developed by \citet{chevallard16} to fit in a Bayesian fashion the stellar population models of \citet{gutkin16} to the observed photometry from U band to IRAC channel 2, fixing the redshift to the systemic value derived in Section \ref{systemic_redshift_estimation}.
We adopt an exponentially delayed star-formation history (SFH $\sim \tau \exp^{-\tau}$, where $\tau$ is the typical star-formation time), a Chabrier IMF, and we calculate the current SFR value over the last $100$ Myr. The value of $\log \tau$ was initialised with a uniform prior ranging $7.0$-$10.5$ (in yr), while the specific SFR (SSFR) and the maximum stellar age were allowed to assume any value in the range $-9.5 < \log(SSFR) < -7.05$ (in yr$^{-1}$) and $7 < \log(age) < 10.2$ (in yr). Similarly, the stellar mass M$_\star$ and the stellar metallicity Z$_\star$ were also assigned, respectively, a uniform prior in the range $5 < \log (M_\star/M_\odot) < 12$ and a Gaussian prior centred at $\log (Z_\star/Z_\odot) = -0.85$ ($1 \sigma=0.2$), following previous results based on the VANDELS survey from \citet{cullen19} and \citet{calabro21}.
The dust attenuation of the galaxy, modelled with a \citet{charlot00} law, is described through the V-band optical attenuation depth $\tau_V$, and initialised with a uniform prior in the range of $0$-$10$ mag. The \citet{inoue14} model is used instead for the attenuation of the intergalactic medium (IGM). 
In our Beagle run, in addition to photometric data, we also fit the observed \xCIII, \xHeII, and \xCIV emission line fluxes (which are the brightest emission features in the far-UV), setting a $2\sigma$ upper limit in case of non-detections. These lines are used by the code to constrain the ionisation parameter (initialised with a uniform prior in the range $-4 < \log (U) < -1$) using the grid of photoionisation models by \citet{gutkin16}. 
This way, Beagle also better takes into account nebular contamination to the photometric bands. 
Finally, we set the dust-to-metal ratio $ \xi_{d}$ and the ratio between interstellar medium depth and total optical depth $\mu = \tau^{ISM}/ \tau_V$ to $0.3$, which are typical values for star-forming galaxies according to \citet{camps16} and \citet{battisti20}.

Plotting together the SFR and stellar mass \mass obtained from Beagle, we see that our subset of $330$ galaxies is representative of the star-forming main sequence at the redshift of this work (Fig. \ref{diagram_Mass_SFR}) in the stellar mass range from $10^8$ to $10^{10}$ \msun, where our median relation is just slightly offset upwards compared to that derived by \citet{speagle14}, but still within their $1 \sigma$ scatter of $0.2$ dex, which is also the average uncertainty of our SFR measurements. For \mass $>10^{10}$, we probe slightly higher specific SFRs, placing galaxies $\sim +0.3$ dex above the MS at $z=3$. 
In this high-mass regime, we also find a small subset of ten galaxies in the starburst regime with SFR at least four times higher than the MS, following the definition by \citet{rodighiero11}.     
As shown by \citet{llerena22}, \CIII emitters are fairly representative of the entire star-forming galaxy population observed with VANDELS (including also non emitters), sharing a similar distribution of mass and SFR. The bias towards higher SFRs at \mass $>10^{10}$ M$_\odot$ is indeed related to the original VANDELS selection, picking up slightly brighter and more star-forming galaxies in that mass range. However, we checked that the results of this paper do not change significantly, even when considering the population of galaxies with \mass $<10^{10}$ M$_\odot$.

\subsection{Morphological analysis}\label{chapter_morphology}

\subsubsection{Merger identification}\label{chapter_merger}

\begin{figure}[t!]
    \centering
    \includegraphics[angle=0,width=1\linewidth,trim={1cm 4.5cm 10cm 0.2cm},clip]{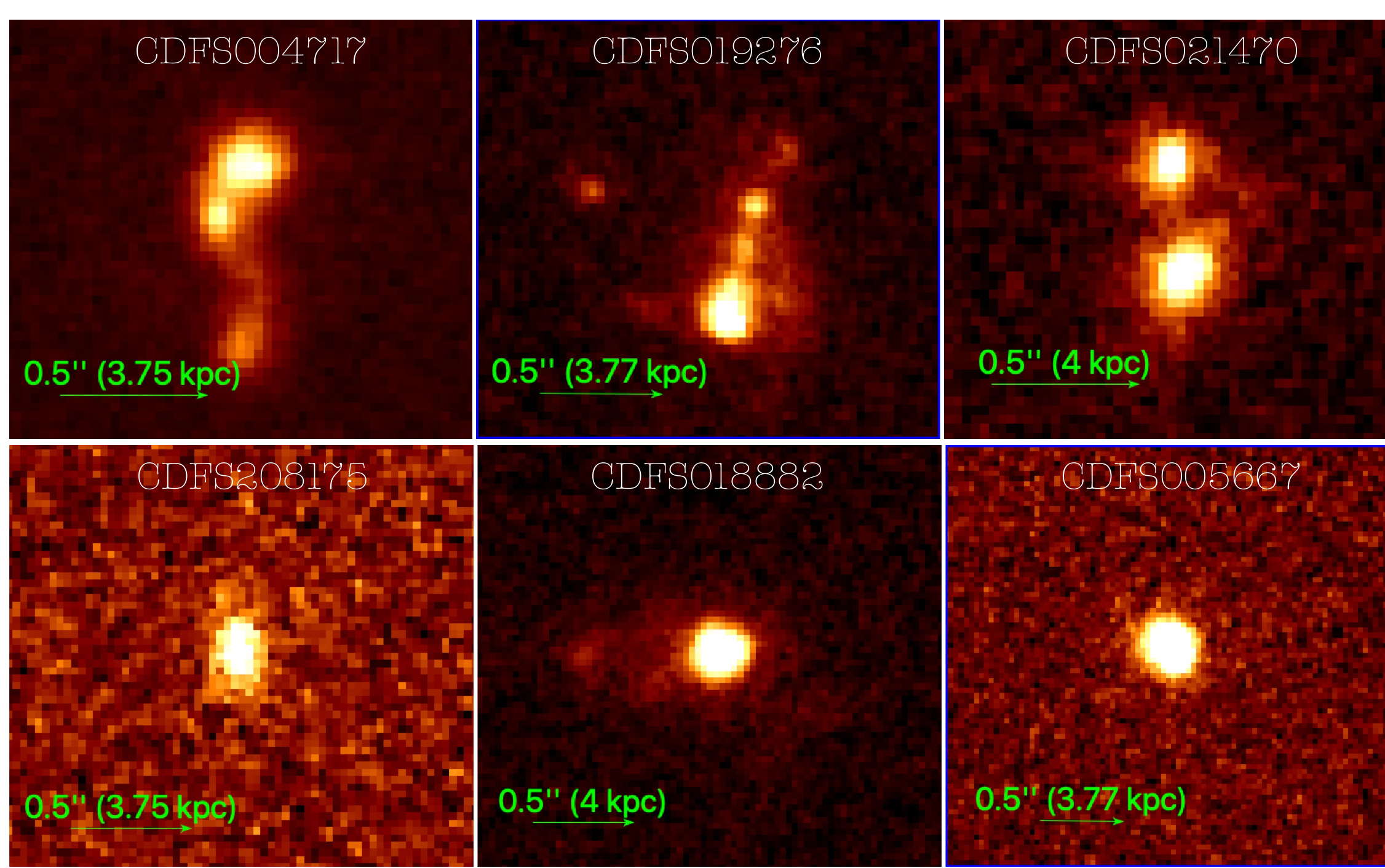}
    \caption{\small HST F814W images \citep{koekemoer11} of three galaxies with reliable systemic redshift and ISM-shift estimations. \textit{Upper row:} These galaxies are visually identified as merging or interacting systems, as showing, from left to right, bridges connecting the interacting components, a very disturbed and asymmetric morphology, and a double nucleus. \textit{Bottom row:} Three examples of isolated or non-interacting galaxies with a discy shape (first case) and a more spheroidal structure (second and third image). 
    }\label{mergers_example}
\end{figure}

Gas inflows or outflows can be produced by galaxy interactions, which can funnel significant fractions of gas towards the centre of the gravitational potential well, usually corresponding to a starburst core \citep{dimatteo05,calabro19} or eject gas into the outskirts as a consequence of tidal forces and feedback \citep{rupke05,dimatteo07}. It is therefore important to carry out a qualitative assessment of the morphology of galaxies, identifying those systems that are undergoing an interaction and where the merger phenomenon can play a role in the gas kinematics. 

A subset of $226$ galaxies from our original sample of \CIII+\HeII emitters have high-resolution HST-ACS F814W images available (probing around $2000$ \AA\ rest-frame at our redshifts), which allows us to perform a statistical morphological analysis and to identify merging systems, with the best resolution and depth achievable with current capabilities. This subset covers not only the CANDELS regions in CDFS and UDS, but is distributed over a wider area, including the wider CDFS field targeted by VANDELS, which in recent years has been covered with HST in the same band $i$. 
As a result, almost all of the originally selected galaxies lying in CDFS and half of those falling in UDS can be used for our morphological analysis. We assigned a merger class to them ($0=$ non-merger, and $1=$ merger) as we explain in the following. 

Given the relatively small size of our sample of \CIII emitters, we decided to identify mergers from visual inspection, which allows us to check interactively and more carefully for the presence of double nuclei and interacting signatures. 
A visual classification has already been applied to the CANDELS fields by \citet[][hereafter K15]{kartaltepe15} for $50000$ galaxies spanning the redshift range $0 < z < 4$, with classifications from three  to five independent people for each object. 
 For their merger sample (i.e. our merger class
$=1$), these authors considered  sources with double nuclei or with evidence of bright tidal features such as tails, loops, bridges, and very disturbed morphologies, all of which are suggestive of an ongoing major interaction. The interaction can be within the same segmentation map, or rather with a companion galaxy that can be still clearly distinguishable. All these cases fall in the interaction classes $1$, $2$, or $3$ in K15. We note that disturbed morphologies in the UV rest-frame can also be due to the presence of bright stellar clumps in a more regular galaxy.

We adopt the K15 classification for galaxies in the CANDELS regions, while we use the same procedure to classify in $i$-band the remaining galaxies that fall in the CDFS wide area targeted by VANDELS. 
In our redshift range, we expect to have an intense growth of galaxies through gas accretion or merging events with smaller companions or satellites, resulting in large gas fractions, unstable discs, and very asymmetric and clumpy morphologies. For this reason, a very conservative approach is adopted in the classification, not including in the merger sample those galaxies with only minor morphological disturbances, such as faint asymmetric features, small companions, or off-centre clumps.

We find a total of $64$ mergers, which represent a fraction of $\sim 30 \%$ of the sample with HST images available. 
The first row of Fig. \ref{mergers_example} shows three examples of merger systems within our \CIII+\HeII line emitters, with different characteristic interacting features. The bottom row also shows the comparison with a subset of different types of isolated, non-interacting galaxies, including both discy and more spheroidal systems. The second of these images explains our conservative procedure: despite the faint asymmetric emission leftward of the main disc, we do not include that galaxy in the merger sample as there is no clear evidence of a major, ongoing interaction. 

\subsubsection{Size measurement and concentration parameter}\label{size_chapter}


\begin{figure}[t!]
    \centering
    \includegraphics[angle=0,width=0.8\linewidth,trim={0.1cm 0.3cm 0cm 0.1cm},clip]{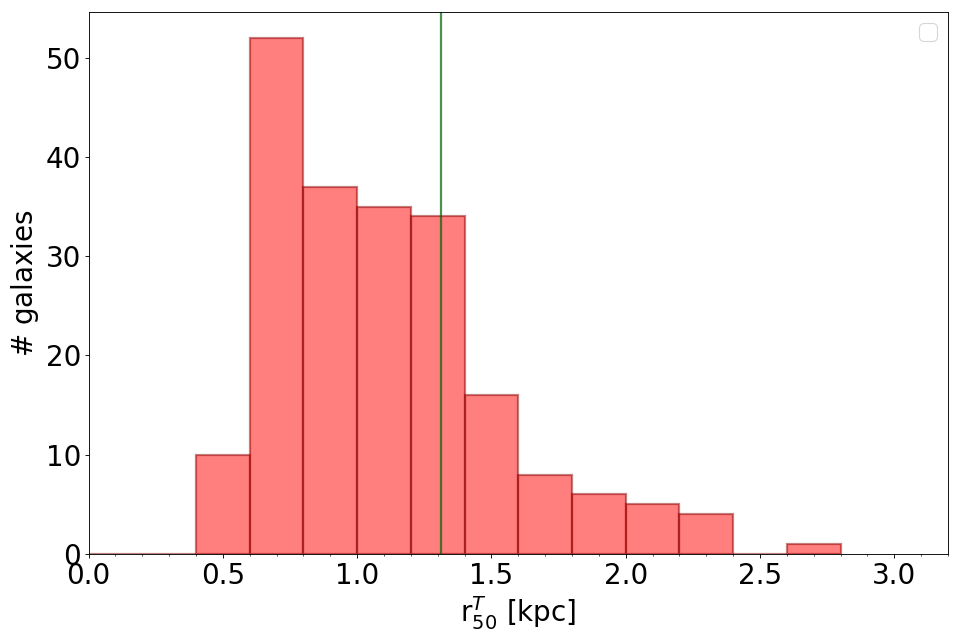}
    \includegraphics[angle=0,width=0.8\linewidth,trim={0.1cm 0.3cm 0.5cm 0.1cm},clip]{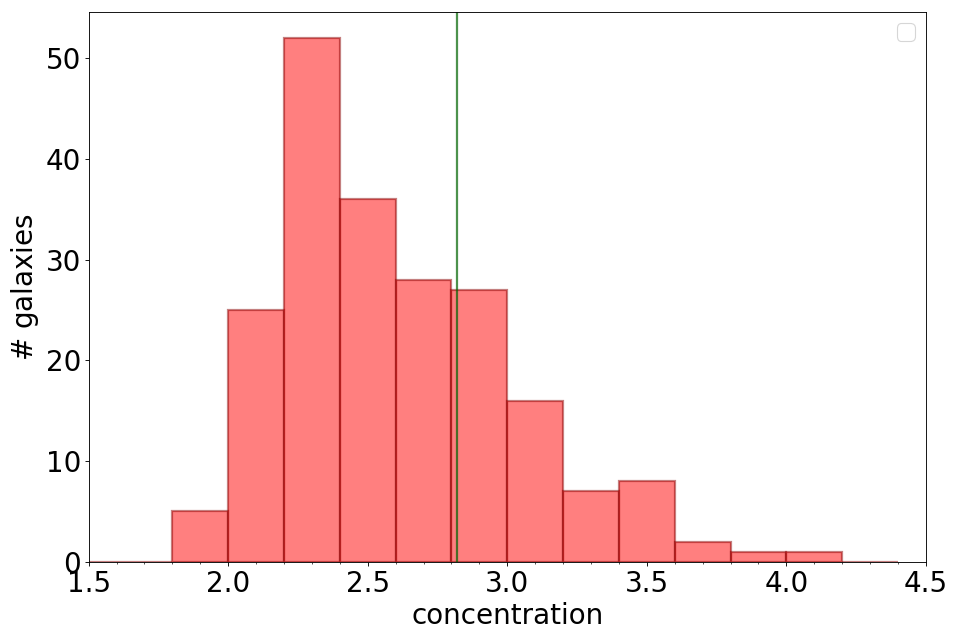}
    \caption{\small Histogram distribution of the equivalent radius \rt and the light concentration parameter \ct (respectively, top and bottom diagram) for the sample of \CIII+\HeII emitters with estimated morphological parameters from the available HST-ACS F814W images. 
    }\label{size_concentration_histograms}
\end{figure}

\begin{figure*}[t!]
    \centering 
    \includegraphics[angle=0,width=0.99\linewidth,trim={0.2cm 0.5cm 1.5cm 0.5cm},clip]{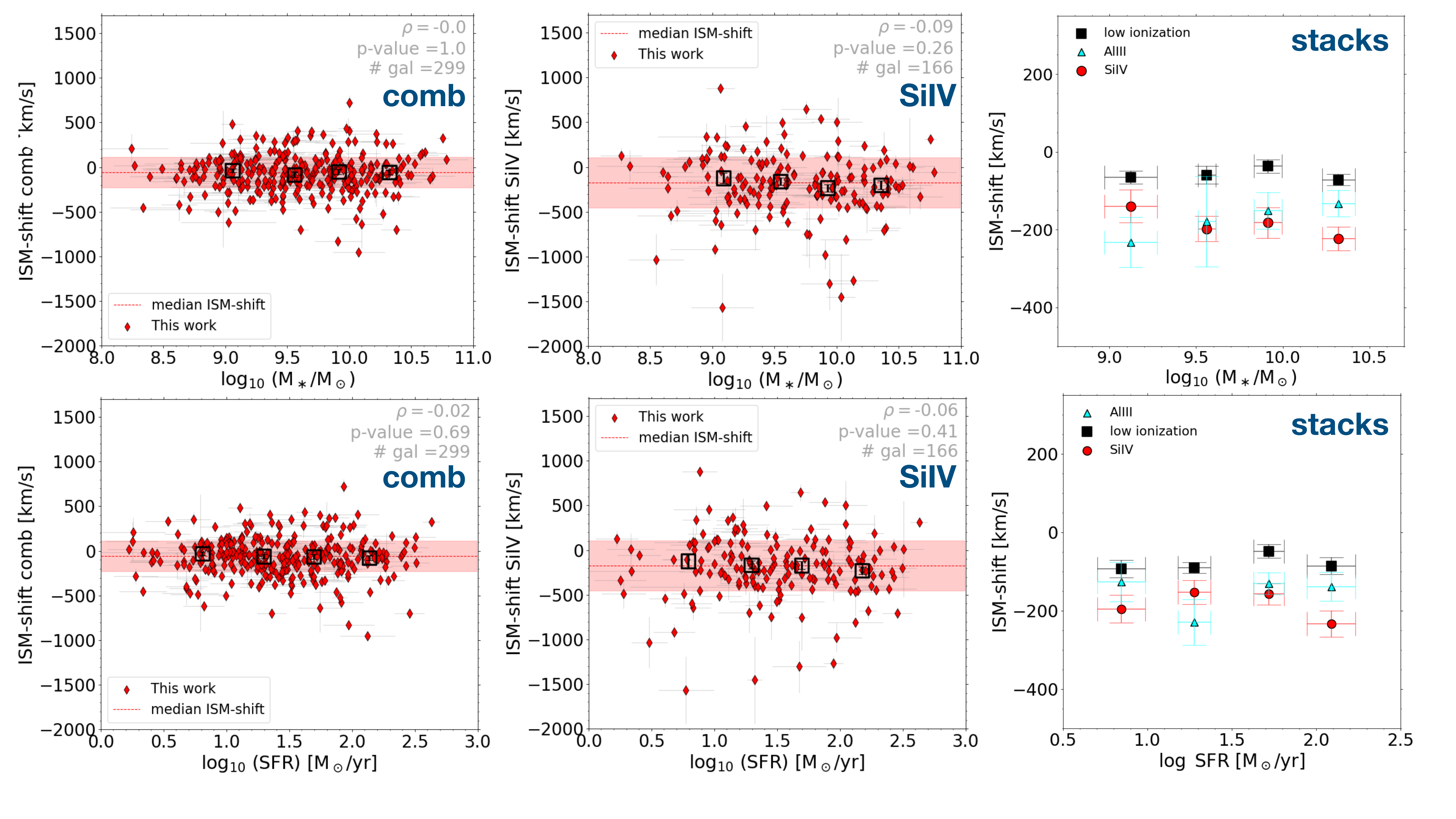}
    \caption{\small \textit{Upper panel:} Correlation of \vismcomb, \vismSiIV and \vismAlIII with the stellar mass \mass for VANDELS star-forming galaxies selected in this work. The first two panels in each row show the relation for individual galaxies (red diamonds), while the black empty squares are the median \vism in bins of increasing \mass. The red horizontal continuous line shows the median \vism for the entire sample shown in each plot, while the red shaded regions highlight the standard deviation of all the ISM shift values. The last panel of the row shows the velocity shifts calculated directly from the spectral stacks in four bins of stellar mass. \textit{Lower panel:} Same as above but as a function of the SFR. 
    }\label{relations_individual_1}
\end{figure*}

\begin{figure*}[t!]
    \centering
    \includegraphics[angle=0,width=0.99\linewidth,trim={0.3cm 2cm 0.2cm 0.1cm},clip]{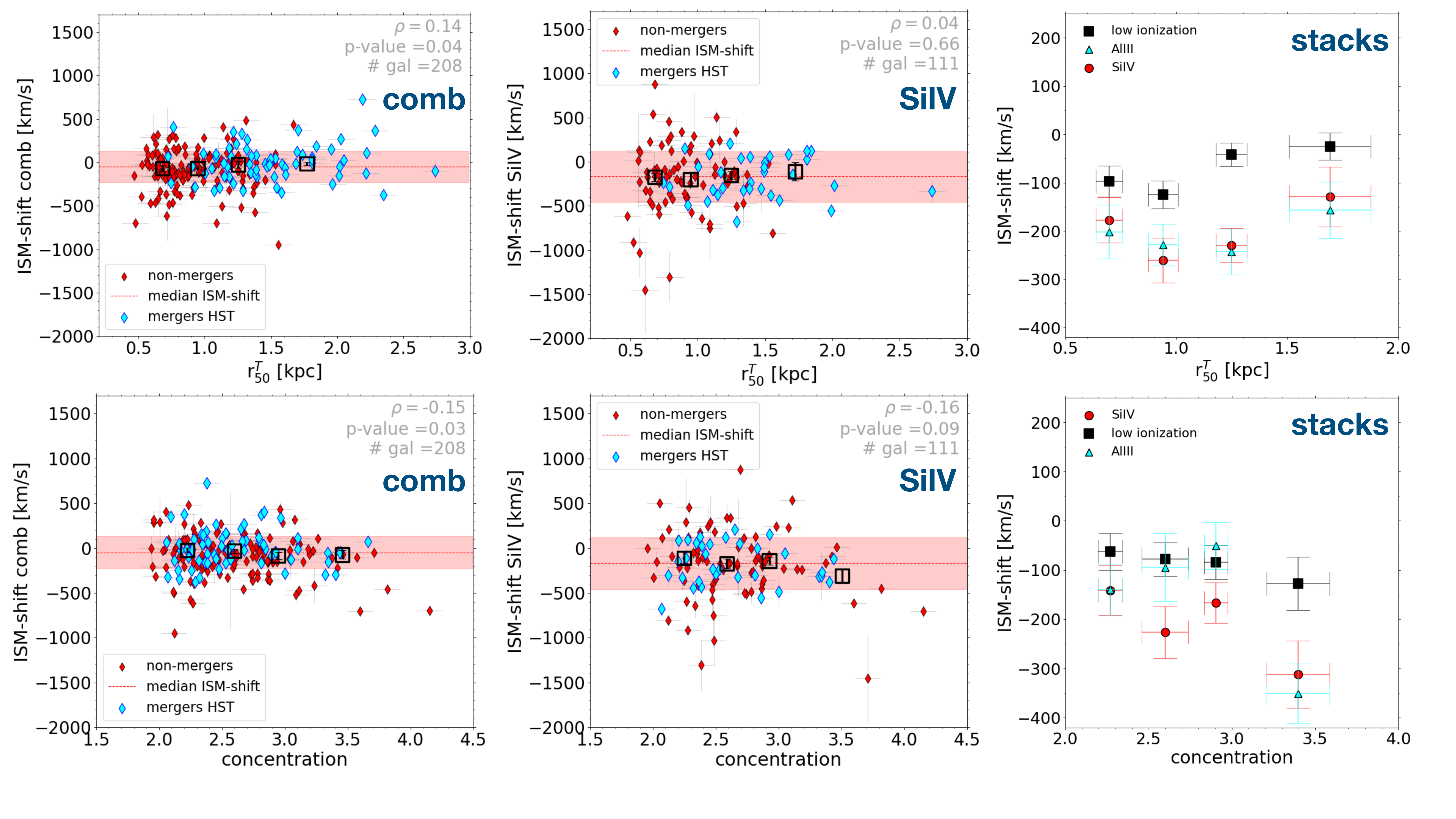}
    \vspace{-0.2cm}
    \includegraphics[angle=0,width=0.99\linewidth,trim={0.3cm 19cm 0.2cm 0.1cm},clip]{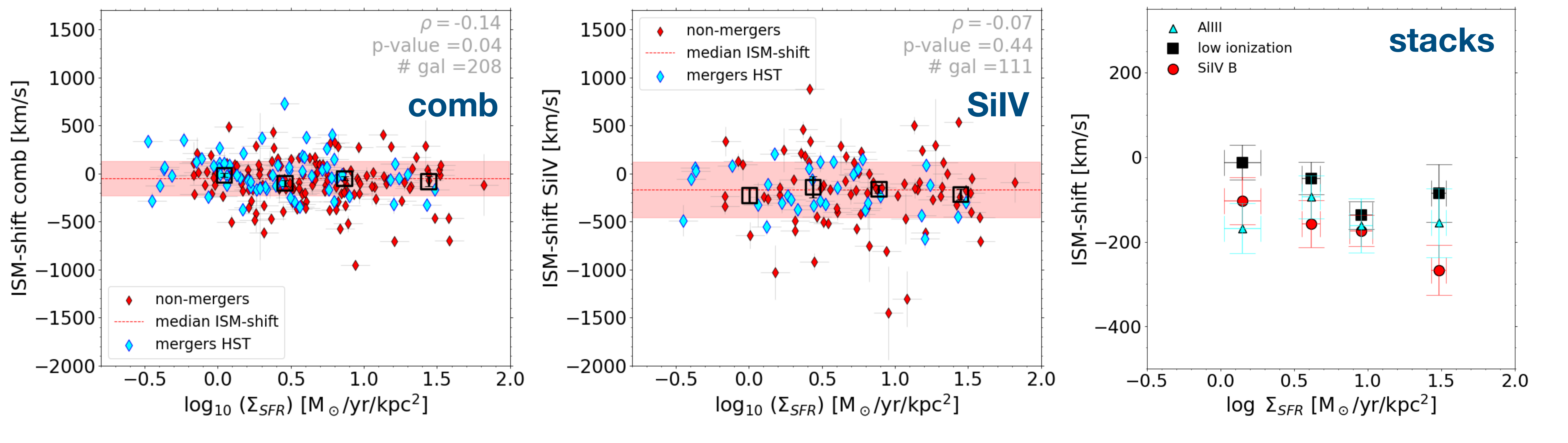}
    \vspace{-0.2cm}
    \caption{\small Diagrams comparing \vismcomb, \vismSiIV and \vismAlIII to the equivalent radius \rt, the concentration parameter \ct, and the SFR surface density \sigmasfr, respectively, from top to bottom row. The derivation of the plots in each single row and the symbols are the same as explained in Fig. \ref{relations_individual_1}. The scatter plots presented in this figure are limited to galaxies with an available HST F814W image.
    }\label{relations_individual_2}
\end{figure*}

\begin{figure}[h!]
    \centering
    \includegraphics[angle=0,width=0.99\linewidth,trim={0.3cm 0cm 0.2cm 0.1cm},clip]{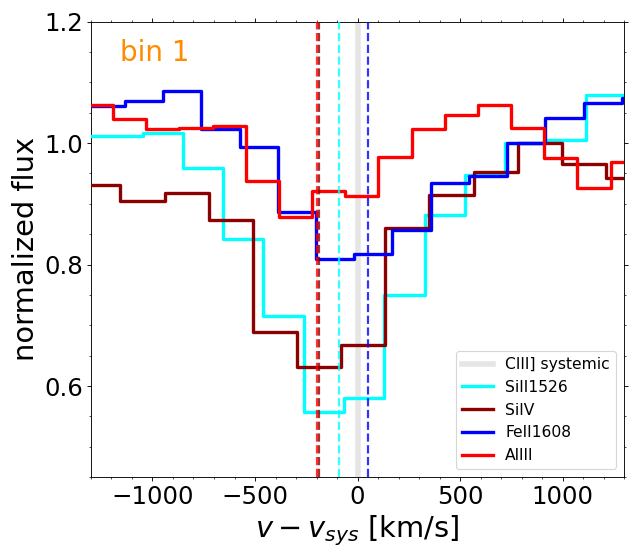}
    \caption{\small Comparison of the line profiles in the spectral stack including all star-forming galaxies selected in this work (AGNs excluded). The central wavelengths resulting from a Gaussian fit are highlighted in cyan, blue, red, and dark red for the \xSiIIb line (representative of low-ionisation lines), \xFeII, \xAlIII, and \xSiIV, respectively. 
    }\label{line_profiles_SF}
\end{figure}

\begin{figure*}[t!]
    \centering 
    \includegraphics[angle=0,width=0.99\linewidth,trim={0.2cm 0.1cm 3.5cm 0.5cm},clip]{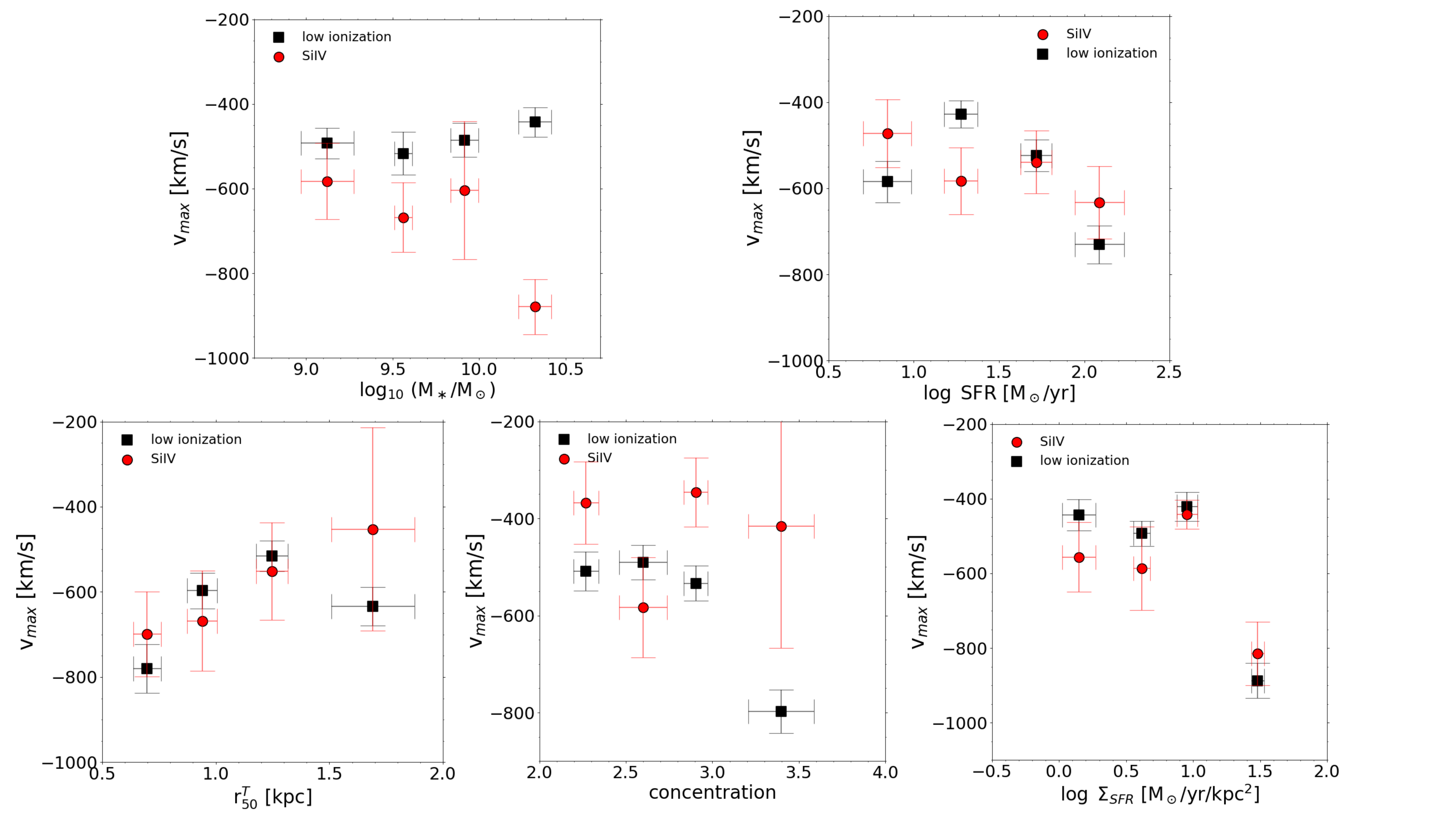}
    \caption{\small Diagrams comparing \vmaxcomb and \vmaxSiIV (estimated from stacked spectra) to other physical properties of the galaxies, namely the stellar mass \mass, the SFR, the equivalent radius \rt, the concentration parameter \ct, and the \sigmasfr.  
    }\label{relations_individual_3}
\end{figure*}

\begin{figure}[t!]
    \centering
    \includegraphics[angle=0,width=0.99\linewidth,trim={0.cm 6cm 39cm 0.1cm},clip]{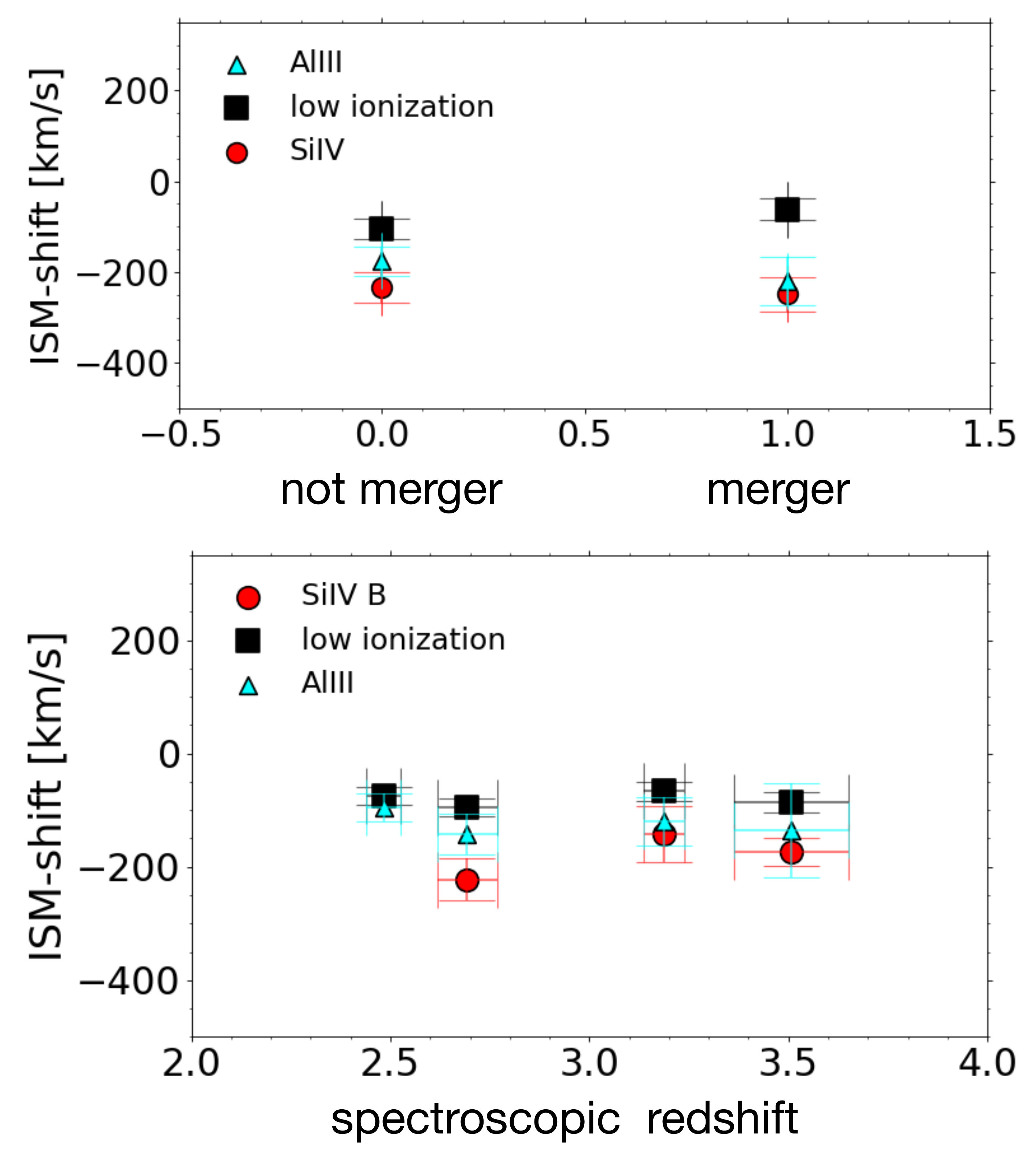}
    \caption{\small Dependence of the ISM velocity shift \vism on the merger flag ($1=$ mergers and $0=$ non-mergers) and spectroscopic redshift. Each symbol represents a spectral stack in a different redshift or merger flag bin. The colours highlight the results from the combined fit of low-ionisation lines (black squares), \SiIV (red circles), and the \AlIII absorption line (cyan triangles). 
    }\label{relations_stacking_comb2}
\end{figure}

In low-redshift observations and simulations, ISM outflows induced by star-formation feedback have been shown to be ubiquitous in galaxies above some SFR surface density (\sigmasfr) threshold, typically $\Sigma_{SFR}> 0.1$ \msun/yr/kpc$^2$ \citep[e.g.][]{heckman11,sharma17}. Several studies also found a correlation of this quantity with the outflow velocity \citep[e.g.][]{heckman15}. It is therefore worth checking for possible correlations between the ISM kinematic properties and \sigmasfr in our work. This parameter, in addition to the SFR, requires measurement of the physical sizes of the star-forming regions in the system. Given that our SFRs are UV-based, we also measure the sizes from optical (observed) HST images, which probe the UV rest frame of the galaxies. 

Because of the large fraction of irregular, asymmetric, and clumpy galaxies at redshifts $\geq 2$ \citep{guo12,huertascompany16}, whose structures do not yet resemble the regular shapes of the Hubble sequence seen at lower redshift, a parametric fitting (e.g. with a Sersic profile) tends to severely underestimate the size \citep[][hereafter R16]{law07,ribeiro16}. Therefore, despite the fact that Sersic parameters are  publicly
available for  the CANDELS data, as provided   by \citet{griffith12} and \citet{vanderwel14}, although these catalogues do not cover the wider CDFS field targeted by VANDELS, we decided to follow a non-parametric approach by adopting the equivalent radius  as a size measurement, using exactly the same procedure defined in R16 and already applied successfully in the VUDS survey at similar redshifts. 
The equivalent radius r$_{T}^{x}$ (in kpc) is defined analytically as:
\begin{equation}\label{radius1}
r_{T}^{x}\ [kpc] = \sqrt{\frac{T_{x}}{\pi} }.
\end{equation}

The variable T$_x$ (in kpc$^2$) in the above equation is given by:
\begin{equation}
  T_x\ [kpc^2]= N_x L^2 \left( 2\times10^{-11} \frac{\mathrm{ster}}{\mathrm{arcsec^{2}}}\right) D_{A}^{2},
  \label{eq:npsize}
\end{equation}
where $N_x$ is the number of pixels (with size $L$ in arcsec/pixel) that sum up to $x \%$ of the total flux of the galaxy, while $D_A$ is the angular diameter distance. The quantity r$_{T}^{x}$ is also corrected for PSF broadening effects following Equation~18 of R16. 

The advantages of this definition are that we do not need to assume a surface brightness profile, and therefore it does not depend on the galaxy shape, on the aperture definitions, on initial guesses, or on the convergence of a fit. 
Even though it can be calculated in principle with any percentage $x$ comprised between $0\%$ and $100\%$, we consider in our work \rt, which counts the brightest pixels summing up to $50\%$ of the galaxy flux, and is more similar to the effective radius  from GALFIT that is often adopted. 

Another useful morphological parameter describing the light profile of the galaxy is the concentration, which was modified from the original definition of \citet{bershady00} to take into account the asymmetric and irregular structure of high-redshift galaxies. The new concentration parameter C$_T$ is calculated (following R16) from the two equivalent radii r$_{T}^{20}$ and r$_{T}^{80}$ as 
\begin{equation}
C_T = 5\times \log_{10}\left( \frac{r_{80}^{T}}{r_{20}^{T}} \right).
\label{equation_concentration}
\end{equation}

Compared to the standard definition of concentration, the advantage of this new quantity is that it does not depend on the definition of a galaxy centre, and it is not affected by the presence of multiple bright clumps or irregular morphologies and therefore works more efficiently in characterising the light profiles in the case of typical galaxies found in the early Universe. As shown by R16, the usual definition tends to underestimate the true light concentration in the case of asymmetric profiles or highly inhomogeneous emission. More details on the derivation of \rt and \ct can be found in R16.

In Fig. \ref{size_concentration_histograms}, we show the histogram distribution of \rt and \ct for our sample of \CIII and \HeII emitters with available HST F814W images and morphological parameters. 
Our galaxies have \rt ranging from $0.3$ to $2.7$ kpc, with the mergers identified in the previous section having amongst the highest spatial extensions, because for these systems \rt includes the size of two merging systems rather than one. 
On the other hand, the light concentration \ct ranges between $2$ and $3.75$, with a median value of $2.5$, which is consistent with the typical values found by R16 (their Fig. 18) for star-forming galaxies from z$=2$ to $z=4$. As discussed in R16, we also find that \rt correlates well with the Sersic radius for a subset of CANDELS galaxies where r$_e$ was estimated in previous works.

\section{Results}\label{results}

In this section, we show how the bulk ISM velocity shift \vism and the maximum velocity \vmax are related to other fundamental galaxy properties that we derived in Section~2. Our aim here is to understand the physical origin of the broad range of \vism and \vmax values spanned by our galaxies. We perform the analysis on \vism both for individual galaxies, where we can check for peculiar ISM velocities, and on a global perspective, using the composite spectra of objects with similar properties. In order to explore the relation globally, we bin the sample in multiple groups, adopting the first, second, and third interquartile values as bin separations, which usually ensures an adequate S/N to our purposes for the stacked spectrum in each bin. 

\subsection{Dependence of the bulk ISM velocity on galaxy properties}\label{results1}

\begin{table*}[t!]

\centering{\textcolor{blue}{Correlation results}}
\renewcommand{\arraystretch}{1.5}  
\vspace{-0.2cm}
\small
\begin{center} { 
\begin{tabular}{ | m{2.4cm} | m{2.4cm}| m{2.4cm} | m{2.4cm} | m{2.4cm}| m{2.4cm} |} 
  \hline
  \textbf{velocity [km/s]} & \textbf{\mass [\msun]} & \textbf{SFR [\msun/yr]} & \rt [kpc] & \ct & \sigmasfr [\msun/yr/kpc$^2$] \\
  \hline
  \vismcomb & \multirow{2}{*}{\parbox[t]{\hsize}{$\rho=0$ ($1$), $m=10\pm30$} } & \multirow{2}{*}{\parbox[t]{\hsize}{$\rho=-0.02$ ($0.69$), $m=-10\pm30$} } & \multirow{2}{*}{\parbox[t]{\hsize}{$\rho=0.14$ ($0.04$), $m=80\pm40$} } & \multirow{2}{*}{\parbox[t]{\hsize}{$\rho=-0.15$ ($0.03$), $m=-80\pm40$} } & \multirow{2}{*}{\parbox[t]{\hsize}{$\rho=-0.14$ ($0.04$), $m=-65\pm30$} } \\
   & & & & & \\
  \hline
  \vismSiIV & \multirow{2}{*}{\parbox[t]{\hsize}{$\rho=-0.09$ ($0.26$), $m=-40\pm50$} } & \multirow{2}{*}{\parbox[t]{\hsize}{$\rho=-0.06$ ($0.41$), $m=-10\pm50$} } & \multirow{2}{*}{\parbox[t]{\hsize}{$\rho=0.04$ ($0.66$), $m=40\pm90$} } & \multirow{2}{*}{\parbox[t]{\hsize}{$\rho=-0.16$ ($0.09$), $m=-130\pm80$} } & \multirow{2}{*}{\parbox[t]{\hsize}{$\rho=-0.07$ ($0.44$), $m=-50\pm60$} } \\
   & & & & & \\
  \hline
\end{tabular} }
\end{center}
\vspace{-0.1cm}
\caption{\small Table indicating the Pearson correlation coefficients $\rho$ (with corresponding p-values) between the ISM velocity shifts of low-ionisation lines and \SiIV, and different physical properties of the galaxies that we analyse in this paper. We also include the slope $m$ of the best-fit line and its uncertainty for each couple of parameters. 
}
\label{table_correlations}
\end{table*}


Following the previous findings, we first test the relations of \vism with the stellar mass and SFR. In particular, we analyse the velocity shift \vismcomb from the combined fit of low-ionisation lines (i.e. \SiIIa, \SiIIb, \CII, and \AlII), and that derived from higher ionisation lines as \vismSiIV. \AlIII behaves similarly to \SiIV but is available for fewer individual objects because it is a fainter absorption, and therefore we consider it later in the stacking analysis. 

The results for \vismcomb and \vismSiIV are shown in Fig. \ref{relations_individual_1} for each of the above two physical parameters. 
In all four diagrams, we do not find any significant correlations between \mass (or SFR) and \vism for individual galaxies. Considering all the sources, we indeed obtain very low Pearson correlation coefficients ($\rho < 0.1$ in absolute value) with high p-values, indicating that our findings are consistent with no dependence between these quantities in the parameter space spanned by our galaxies.
We also divide our sample in four subsets using the interquartile ranges of the stellar mass and SFR, calculating the median \mass and SFR of all the galaxies residing in the same bin, and then estimating \vismcomb and \vismSiIV from the stacks. This way, we obtain a median relation representative of the whole population, drawn with black empty squares in Fig. \ref{relations_individual_1}, which also suggests that the trend is not significantly different from constant for both low-ionisation absorption lines and for \SiIV.
As confirmation of this result, the angular coefficient of the best-fit lines to all the individual points in the diagrams is always consistent with zero. 
We can also better appreciate the larger dispersion of \SiIV velocities compared to the lower ionisation lines, as we have seen in Fig. \ref{histograms_ISMshift}, and in particular the higher number of sources with ISM velocities $< -500$ km/s, that is, higher outflow velocity.   
Interestingly, the presence of outliers or peculiar systems, with inflow (outflow) velocities above (below) the median velocity by more than $1\sigma$, does not correlate with either \mass or SFR in any of the cases. These extreme cases therefore do not seem to have different properties compared to the average population.

In the last column of Fig. \ref{relations_individual_1}, we show the results obtained in a different way by measuring the ISM-velocity shifts directly from the spectral stacks instead of individual galaxies. 
In this case, we also add the measurements for the \xAlIII doublet, which is better detected in the stacked spectra. Using the same previous bins of \mass and SFR, we can understand from this exercise that also from a global perspective there are no significant variations as a function of these two quantities for \vismcomb, \vismSiIV, and \vismAlIII.    

We then analysed the dependence of the ISM velocity shift on morphology-related parameters, namely the physical size as \rt, the concentration \ct, and the SFR surface density \sigmasfr. 
The results are shown in Fig. \ref{relations_individual_2} for both individual galaxies and for the spectral stacks, after defining bins of increasing \rt, \ct, and \sigmasfr. As before, we focus on the velocity shifts of the LIS (through the combined fit) and of higher ionisation lines including \AlIII and \SiIV. 

In the first row, where physical size is represented, we find a weak and marginally significant correlation between \rt and \vism, as indicated by the low Pearson correlation coefficient (and p-value $<0.05$) from both the combined fit of LIS lines. We should note that the median \vism is slightly lower (i.e. higher outflow velocity) for galaxies of smaller sizes, and this is also true when considering the \SiIV line.  
In the stacking analysis in the last panel, \vismcomb of galaxies with \rt $\gtrsim1.2$ kpc is below the median population value at the $1\sigma$ level, and its difference compared to smaller galaxies is significant at $2\sigma$. A similar difference is also found for \AlIII and \SiIV between galaxies with \rt lower and higher than $\sim 1.5$ kpc. 

For the concentration parameter displayed in the second row of Fig. \ref{relations_individual_2}, we can see that a correlation, although weak, exists with \vismcomb and \vismSiIV already for individual galaxies, as indicated by the results of the Pearson correlation test. 
Moreover, we can see in the third panel that the stack of galaxies with a higher concentration parameter (\ct $>3$) has \vismcomb, \vismSiIV , and \vismAlIII that are lower by $\sim50$, $100,$ and $150$ km/s, respectively, compared to galaxies with less concentrated UV emission and \ct $\leq 3$, with a significance of $2\sigma$ on average. However, we also notice that the lower \vismcomb and \vismSiIV at high concentration parameter is mostly driven by non-merger systems (Fig. \ref{relations_individual_2}).  This difference is similar to the downward offset that we observe in the median \vism of the \SiIV line for galaxies residing in the last bin of \ct (second panel of the second row). 

Finally, we show the relation with \sigmasfr in the last row of Fig. \ref{relations_individual_2}. Given the weak correlation with \rt, with this new parameter we also find a similarly weak dependence on \vism. Indeed, the Pearson correlation coefficient and the linear best-fit slope indicate a significance of $2\sigma$ when considering the low-ionisation lines for individual galaxies. For \SiIV, even though the correlation is not significant, we notice that the highest outflow velocities (i.e. lowest \vismSiIV) also tend to have higher \sigmasfr  in the stacked analysis.

Overall, there are no significant correlations between the bulk ISM velocity (in all the ionisation states that we probed) on stellar mass and SFR, while a weak trend (marginally significant at $2$ $\sigma$) is found for the other three parameters, indicating that galaxies with more concentrated emission (\ct $\gtrsim 3$), smaller size, and higher \sigmasfr might launch faster outflows. 

As already shown in Fig.~4, in all the right panels of Figs. \ref{relations_individual_1} and \ref{relations_individual_2}, we then notice on average a systematic difference in \vism among the various types of far-UV absorption lines that we detect in VANDELS spectra.
The combined fit, which traces the kinematics of the low-ionisation lines, indicates that the neutral ISM in a typical star-forming galaxy is outflowing, with average $|\vismcomb|$ of $60\pm10$ km/s. On the other hand, higher ionisation lines such as \SiIV and \AlIII trace gas that is moving away from galactic centres at higher velocities of $170\pm30$ km/s and $160\pm30$ km/s, respectively.
In Fig. \ref{line_profiles_SF}, we can see the different velocity profiles normalised to the range $+700$-$1000$ km/s of the absorption lines in the stacked spectrum constructed from the full sample of $330$ star-forming galaxies with \CIII or \HeII emission and therefore with a systemic redshift measurement.
The Gaussian centroid of \xSiIIb that we take as representative of low-ionisation lines tracing neutral gas is shifted towards negative velocities ($-100$ km/s) compared to the systemic redshift. 
\xFeII has a broader profile, is less deep, and is centred at slightly positive ISM shifts close to the systemic redshift. On the other hand, \xSiIV and \xAlIII have blueshifted Gaussian centroids, with larger \vism in outflow by $100$ km/s compared to the low-ionisation line \SiIIb. We further discuss the velocity difference between these two gas phases in Section \ref{lower_higher_ionisation_lines}. 
In Table \ref{table_correlations}, we summarise the correlation strengths between \vism and all the galaxy properties studied in this work.

\subsection{Dependence of the maximum ISM velocity on galaxy properties}\label{results2}

We now explore the dependence of the maximum velocity for the same galaxy parameters seen in the previous section. Compared to the bulk ISM velocity, \vmax carries more information about the gas flows as it also depends on the absorption line width. As mentioned in Section \ref{vmax}, we performed the analysis on the spectral stacks, and we consider only the combined fit of low-ionisation lines and \SiIV for which we have a higher S/N. 

In order to test the dependence with the same physical properties analysed in the previous section, we divided our sample into four groups of similar properties (using the interquartile ranges), and then stacked the spectra and calculated \vmax in each bin. The diagrams for all the parameters are shown in Fig. \ref{relations_individual_3}. 
Here we can see that globally, we find similar results to the previous analysis based on the bulk ISM velocity.
In particular, we find a similar decreasing trend of \vmax for smaller galaxies and highly concentrated systems (significance $>3 \sigma$), although the second case is valid only from the combined fit where we have a higher S/N. Compared to the previous section, we find here a marginally significant ($\sim 2 \sigma$) relation between \vmax and both the SFR and \sigmasfr, for both the lower and higher ionisation lines. In the latter diagram, the weak trend is mainly driven by the last bin, where galaxies with \sigmasfr $>1$ \msun/yr/kpc$^2$ show maximum outflow velocities that are approximately $300$ km/s higher than the remaining population at lower \sigmasfr. Finally, for the stellar mass we also find more negative \vmax at \mass $>10^{10}$ \msun, but only from the \SiIV line analysis. 
Overall, we find that \vmax is on average better correlated with the galaxy physical properties compared to \vism, although the correlations remain weak and marginally significant, and, similarly, slightly more important for morphological related quantities.

\subsection{Outflows in mergers and redshift evolution}\label{results3}

Finally, we also tested the effects of mergers on gas kinematics in high-redshift star-forming galaxies. According to \citet{chisholm15}, mergers at $z\sim0$ can drive faster outflows than isolated galaxies. 
In our case, we can see already in Fig. \ref{relations_individual_1} and \ref{relations_individual_2} for individual galaxies that visually identified mergers (cyan diamonds) are distributed over the whole range of explored parameters and have a similar distribution in \mass, SFR, \sigmasfr , and concentration to that of more isolated galaxies. In the first row of Fig. \ref{relations_individual_2}, we note that mergers tend to have larger sizes \rt, which is reasonable considering that they are often composed of two or more interacting components, each with its own size. 

Furthermore, we have performed a stack of all identified merger systems in the selected sample, in order to better compare with the rest of the population of non-merger galaxies. The results are shown in the top panel of Fig. \ref{relations_stacking_comb2}, in which we can see that mergers and non-merger galaxies (respectively indicated with a merger label $1$ or $0$) have similar ISM kinematics. Indeed, we find that the bulk velocities are consistent within the uncertainties at the $1$-$2\sigma$ level for all the lines, both the higher ionisation (as \vismSiIV and \vismAlIII) and lower ionisation lines (\vismcomb). 

Finally, considering the wide range of redshifts (from $2$ to $5$) probed by VANDELS, as represented in Fig. \ref{redshift_distribution}, we also checked for a possible evolution in redshift of the outflow properties in star-forming galaxies, measuring \vismcomb, \vismSiIV and \vismAlIII in spectral stacks in bins of increasing redshift.
The results of this exercise are shown in the bottom panel of Fig. \ref{relations_stacking_comb2}, and again indicate that there is no difference in the average outflow velocity as a function of redshift for the low and higher ionisation lines, as no significant correlation is found. Despite the slightly higher outflow velocity from \SiIV at $z\sim 2.7$, this is still consistent with the other bins within $1\sigma$.


\section{Discussion}\label{discussion}

We show in the above sections that, considering the galaxy population as a whole, the interstellar medium traced by UV absorption lines is on average moving with a velocity of $-60$ km/s along the line of sight (hence globally in outflow) for the low-ionisation gas, which rises up to $\sim -170$ km/s for higher ionisation gas. This is confirmed both by averaging the ISM velocities of individual galaxies and by performing spectral stacks in bins of different physical parameters. 
Furthermore, we measure maximum velocities of the gas in the outflow direction ranging between $500$ and $800$ km/s for most of our sample. In the following, we compare our results to studies of the ISM kinematics at lower redshift, and then move to a more physical investigation of the effects of outflows on galaxy evolution.

\subsection{Comparison with works at lower redshifts}\label{discussion1}

Galactic outflows at different cosmic epochs have been studied for decades. 
In the local Universe, \citet{heckman15} studied star-formation-driven winds from far-UV absorption lines (\SiIIa,\SiIIb,\CII,and \SiIV) in a sample of $39$ galaxies spanning a broad range of stellar masses and SFRs, detecting bulk ISM velocities from $100$ to $200$ km/s, globally in outflow. Interestingly, they report strong correlations between ISM bulk velocities and both SFR and \sigmasfr. 

At intermediate redshifts, ISM kinematic properties have been characterised in large samples of star-forming galaxies. \citet{bordoloi14} analysed $486$ galaxies at $1 \leq$ z $\leq 1.5$, with $\log$(\mass/\msun) and $\log$(SFR/\msun/yr) ranging from $9.45$ to $10.7$ and $0.14$ to $2.35$, respectively, finding typical ISM velocities \vism of $\sim 150$-$200$ km/s from the MgII absorption doublet. Furthermore, \vism was found to correlate with \sigmasfr but not with the SFR itself. 
Furthermore, \citet{weiner09} used the same MgII absorption to calculate \vism for a sample of $\sim1400$ galaxies at $z\sim1.4$ with similarly massive galaxies of the previously mentioned work, and a broad range of SFRs from $10$ to $100$ \msun/yr. These authors found slightly higher typical outflow velocities of $\sim300$ to $500$ km/s, increasing both as a function of stellar mass and SFR, in analogy to local infrared-luminous galaxies.

If we move to earlier cosmic epochs, \citet{talia12} studied ISM flows with the same methodology as that used here in $74$ star-forming galaxies in the stellar mass range from $10^{9.2}$ to $10^{10.6}$ \msun at redshifts of $1.4$ to $2.8$ (z$_{median}\sim 2$), finding a median ISM velocity of $100$ km/s and no significant dependence of \vism on galaxy properties. 
Similarly, \citet{steidel10} derived \vism from the position of \CII, \SiIV, \SiIIa, and \SiIIb far-UV absorption lines for a statistical sample of $89$ Lyman-break galaxies at $2<z<2.6$ (z$_{median}=2.3$) and slightly more massive (\mass $> 10^{9.5}$ \msun). While these later authors found a slightly higher ISM velocity than us ($=164 \pm 16$ km/s), they report no correlations of \vism with the main galaxy properties including \mass and SFR.
Our results therefore extend this lack of correlation up to at least $z\sim4.5$. 

Interestingly, we find on average ISM velocities closer to zero compared to previous works. Moreover, as noted in Section \ref{galaxies_with_inflows}, we find that $34\%$ of our galaxy sample has a positive velocity shift, indicative of a global inflow. This fraction is approximately double that found by \citet{steidel10}, and we reach higher inflow velocities on average for the low- and high-ionisation absorption lines.  
Overall, our findings suggest that, for the same levels of SFR and stellar mass of the host galaxies, the bulk of the ISM moves closer to systemic from low redshift to high redshift, and becomes less sensitive to variations of the star formation activity and mass budget. 
This might be due to an increased role of inflows at earlier cosmic times, which might prevail over star-formation-driven outflows in the majority of systems. 
Regarding these inflows, while we know from theory that cosmological metal-poor gas condensation and accretion onto galaxies should occur at early epochs ($z\gtrsim 2$), it is difficult from our data and measured velocities to disentangle gas accretion from the cosmic web from that originating in galactic fountains. A key test would be the measurement of the gas-phase metallicity of the outflowing and inflowing gas, which would require a higher spectral resolution and sensitivity, and will be addressed in a future work.
 
We remark that we are exploring an epoch where the bulk of the galaxy population is rapidly assembling through mergers and gas inflows, and they have a more gas-rich and turbulent ISM. 
Combining gas infall episodes and these additional effects to stellar-driven feedback might produce multiple absorption components with different relative velocities and with different absorption strengths, of which we measure only a median, spatially integrated, line-of-sight-dependent value for each galaxy.
In other words, we might have as a net result the weakening of outflow signatures compared to lower redshift star-forming galaxies, even though part of the gas can still be moving outwards with high velocity in some galaxies, with some material reaching high speeds of up to \vism $\sim 1000$ km/s. 

At the same time, the weaknesses and lack of correlation of \vism with the main galaxy properties can be explained by the same arguments. However, additional effects might be at play. For example, \citet{bordoloi14} find that face-on galaxies at intermediate redshift ($1$-$1.5$) have higher velocity outflows as compared to edge-on galaxies at similar redshifts and stellar masses, while \citet{robertsborsani20} find that outflows are most powerful in the central regions of massive star-forming galaxies, even though this was studied in the local Universe. 
In order to check these additional dependences, we need to observe statistical samples with IFU instruments and better spatial resolution in order to disentangle the different components of the ISM in high-redshift galaxies. 
We also note that, because of the slightly biased selection (towards higher SFRs) at high stellar masses (\mass $> 10^{10}$ \msun), the trend is less representative of the global star-forming population in that regime, and completely unconstrained at even higher masses than those probed by VANDELS (i.e. \mass $\gtrsim 10^{11}$ \msun). Future spectroscopic surveys probing larger volumes could help to better constrain this parameter space.

\subsection{Outflow physical properties and mass loading factor}\label{discussion2}

In this section, we focus on the physical properties and driving mechanisms of outflows in the average star-forming population at $z\sim3.5$. Therefore, in the following part of the paper, we consider only VANDELS galaxies for which we measure a global outflow, that is, a negative bulk ISM velocity \vismcomb (or \vismSiIV). If we restrict the sample to these objects, 
we obtain average outflow velocities of $-146 \pm 10$ km/s for the former and $-270 \pm 30$ km/s for the latter, which are more similar to the typical ISM velocities measured at intermediate redshifts in the works described in the previous section. We remark that even if we consider only this reduced sample, there would be no significant changes in the correlation results presented in sections \ref{results1} and \ref{results2} between the bulk outflow velocity (or \vmax) and the other galaxy physical properties.   
Taking these results on the outflow velocities, we are now ready to quantitatively assess the effects of star-formation-driven winds on the host galaxies by computing the mass-outflow rate, mass-loading factor, and typical escape velocities of our systems. 

We calculate the mass-outflow rate \Mout in a thin shell approximation as
\begin{equation}\label{eq_Mout}
\rm{\dot{M}_{out} \;=\; 4\pi \;C_{f} \;C_{\Omega}\; \mu m_{p}\; N_{H} \;R \;v_{ISM},}
\end{equation}

where $C_{\Omega}$ and $C_{f}$ are the angular and clumpiness covering fraction, respectively, $\mu m_{p}$  is the mean atomic weight, $R$ is the radius of the thin shell wind, and $v_{ISM}$ is the gas outflow velocity, assumed to be the typical bulk ISM velocity measured from our spectra. As we are deriving an average value over many galaxies viewed from different angles, and because in the stacking procedure we are integrating over lines of sight and clumpiness with different covering fractions, we fix $C_{\Omega}C_{f}\; =\; 1$, as in \citet{bordoloi14}. We also set $\mu = 1.4$, and we infer the outflow radius R as the average equivalent radius estimated for our sample in Section \ref{size_chapter}, which is $\sim1$ kpc. 
Finally, N$_H$ is the column density of the outflowing gas, which can be estimated as in \citet{leitherer11} and \citet{heckman11} from the dust attenuation properties of our galaxies (assuming a Calzetti law), taking into account also a secondary dependence with the metallicity, as

\begin{equation}\label{eq_NH}
N_H= 3.6 \times 10^{21} \times \frac{E(B-V)}{Z_{gas}/Z_\odot}
.\end{equation}

Taking the median E(B-V)$=0.2$ mag measured for our sample from the available photometric data as described in \citet{calabro21}, and considering the typical gas-phase metallicity of $0.4\ \times$ Z$_\odot$ derived for VANDELS galaxies at $z\sim 3$ \citep{cullen21}, we infer N$_H =$ $1.75\times10^{21}$ $cm^{-2}$. 
Using all these values in Eq. \ref{eq_Mout}, we get

\begin{equation}\label{eq_Mout2}
\rm{\dot{M}_{out} [M_\odot/yr] \;\simeq 36\ \frac{N_H}{10^{21.24}\ cm^{-2}} \times \frac{R}{1\ kpc} \times \frac{v_{ISM}}{146\ km/s}}
.\end{equation}

The mass outflow rate estimated in this way is comparable to the SFR of the galaxies, and yields an outflow mass loading factor $\eta$ of the order of unity, or just slightly above. In particular, considering the median SFR of our sample of $196$ galaxies with a net outflow signature (SFR$_{median} = 27$ \msun/yr), we derive:

\begin{equation}\label{eq_eta}
\rm{\eta = \frac{\dot{M}_{out}}{SFR} \simeq 1.3},
\end{equation}
which is nicely consistent with the values found with a similar methodology for low-ionisation gas outflows in local and intermediate redshift ($z \sim 1$-$1.5$) star-forming galaxies \citep{heckman15,bordoloi14}. We also note that a similar $\eta$ was found for ionised and cold molecular outflows in purely star-forming galaxies at $z<0.2$ \citep{fluetsch19}, once we homogenise the assumptions on the wind geometry adopted in the two papers. However, these later authors probe $1.5$ dex higher stellar masses than our study. 

With the above information, we can study the nature of the outflows.
From a physical perspective, when stellar winds impact the surrounding ISM, a shock front is created, which injects energy into the gaseous medium and then propagates outwards, sweeping the gas away from the galactic centre.
The outflows can be classified into two types, depending on the efficiency of the radiative cooling during the shock propagation. 
In momentum-driven flows, the total momentum is conserved, while most of the preshock kinetic energy is radiatively lost, which produces a rapid cooling of baryons in a thin shell of gas and a strong discontinuity with the post-shock region. 
On the other hand, in energy-driven flows, the shock expands adiabatically into the ISM preserving most of its mechanical luminosity, and therefore cooling is negligible, and the flow is much more violent compared to the previous mechanism. 
In order to identify the nature of the outflow and distinguish between the two driving mechanisms, we can compare the energy and momentum rate of the outflow to the momentum and luminosity input from star formation. 
The outflow-related properties are written, respectively, as $\dot{p}_{out}=\dot{M}_{out}\;v_{ISM}$ and $\dot{E}_{out}=0.5\;\dot{M}_{out}\;v_{ISM}^2$. The momentum and energy from star formation are instead given by $\dot{p}_{rad}=L_{SFR}/c$ and $\dot{E}_{rad}=L_{SFR}$, where $L_{SFR}$/L$_\odot \sim 10^{10} \times$ SFR/ \msun/yr \citep{kennicutt98}. Assuming the median SFR of our sample ($=23$ \msun/yr), this yields

\begin{equation}\label{eq_mom1}
\rm{\frac{\dot{p}_{out}}{\dot{p}_{rad}} \;= 0.2 \;\; ; \;\;}
\rm{\frac{\dot{E}_{out}}{{L}_{SFR}} \;= 2\times 10^{-5}},
\end{equation}
which indicates that we are witnessing momentum-driven winds.

Finally, we study the fate of the outflows in VANDELS galaxies by comparing the observed outflow velocities to those needed to escape from the gravitational potential wells of the galaxies. The escape velocity is related, using dynamical arguments, to the circular velocity $v_c$ as $v_{esc} \simeq 3\times v_c$ for reasonable halo mass distributions \citep{binney87,heckman00}. The circular velocity instead can be estimated observationally from the emission line width $\sigma_{vel}$ along the line of sight using the relation $\sigma_{vel}$ $\sim 0.6 \times v_c$, where $\sigma_{vel}$ represents an average over all the observed inclinations \citep[Fig. 7 of][]{rix97}. 
Putting this information together, we obtain the relation $v_{esc} \simeq 5 \times \sigma_{vel}$. We remark that this is a median relation that is subject to large uncertainties depending on the geometry and internal dynamics of the galaxies; however it gives a useful indication, and can be  compared with our ISM velocities. 

In the far-UV spectral range covered by VANDELS, the brightest emission lines are \CIII and \HeII. While the former is a doublet and difficult to model given the resolution of VANDELS, the latter has in general a lower S/N and a larger scatter, with a skewed distribution towards broader profiles compared to \CIII. For these reasons, we instead used the optical rest-frame \OII line, which was also used for the calculation of $v_{esc}$ in the previously mentioned works. In our case, this line was covered by the follow-up of a subset of galaxies with MOSFIRE, which has a resolving power reaching $\simeq 3650$ in band H \citep{cullen21}.

Considering the values of $11$ individual galaxies where the \OII line was detected with a minimum S/N of $5$, and which have a similar mass distribution to our subset, we obtain an average $\sigma_{[OII]}$ of $125$ km/s. Using this estimate as representative of our sample to calculate the escape fraction, we get $v_{esc}$ of the order of $625$ km/s. 
This is well above the average bulk outflow velocities that we measure in our galaxies from both lower and higher ionisation lines. It is also higher compared to the estimates of \vismcomb and \vismSiIV for the majority of individual galaxies. However, it is intereresting to note that a small fraction of objects, where the bulk outflow velocity is amongst the highest in our sample, might be the best candidates to launch significant amounts of gas outside of the systems. 

Considering the stacking analysis performed in Section \ref{results2}, we note that even the average maximum velocity \vmax, as defined in Section \ref{extra_outflows}, does not generally overcome $v_{esc}$. This implies that, for the average population and for a broad variety of conditions, most of the ISM is expected to remain bound to the main galaxy halo, even though we might exceed this limit when we consider galaxies with the highest stellar masses, SFRs, \sigmasfr, and concentration (i.e. smaller size) among those probed in this work. Therefore, there is a greater probability that gas will escape from galaxies with more extreme properties.

\subsection{Low- versus high-ionisation gas outflows}\label{lower_higher_ionisation_lines}

We find that the ISM-shift of \SiIV and \AlIII indicates a faster velocity of the higher ionisation outflowing gas by approximately $100$ km/s compared to \vismcomb of low-ionisation lines (i.e. \SiII, \CII, \AlII) derived from the combined fit. To understand the physical origin of this difference of bulk outflow velocities (and also \vmax), we need to better investigate the properties of our absorption lines. 

A first approach is based on looking at the ionisation potential required for the creation of a particular species. The \SiIV and \AlIII species require an ionisation potential of 45 eV and 28 eV, respectively, which is higher than low-ionisation lines, which need ionisation energies of 16, 19, and 24 eV for the \SiII, \AlII, and \CII ions, respectively. It is clear that, by their nature, low-ionisation elements require denser, self-shielded gas to survive while the high-ionisation elements can survive in more rarefied, less dense regions.
Secondly, we can use different transitions of the same ionisation species to probe the optical depth of the absorbing gas. In the case of high-ionisation lines, they come already in close doublets, while for the low-ionisation gas, we can use the two transitions of the \SiII ion. The individual components of the high-ionisation doublets have similar oscillator strengths $f_1$ and $f_2$ of $\sim 0.5$ and $0.25$ for the bluer and redder component, respectively, indicating that in both cases we should expect an equivalent width ratio of EW$_1$/EW$_2$ = ( $\lambda_1^2 \times f_1$) / ($\lambda_2^2 \times f_2$) $= 0.5$ in the optically thin case \citep[see also][]{erb12}. 
For the \SiIV doublet, we observe an EW ratio of $0.67_{0.6}^{0.75}$, while for \AlIII we obtain a similar value of $0.67_{0.57}^{0.76}$. In an optically thick absorbing gas, the ratio no longer depends on the oscillator strength and approaches a value of $1$. The results imply that we are in an intermediate condition, even though we are actually closer to the optically thin limit. 
For the two transitions of the \SiII ion, at $\simeq 1260$ \AA\ and $1526$ \AA, their different oscillator strengths ($f_1 = 1.22$ and $f_2 = 0.13$) imply an expected ratio in the optically thin case of $0.156$.
We measure an observed ratio of $1.4_{0.5}^{3.5}$, where the larger uncertainties derive from the fact that the two lines are not close in wavelength and do not share the same continuum level as in the high-ionisation doublets. Despite the large uncertainty, this calculation indicates that the EW ratio is consistent with $1$ and well above the optically thin value. 

Both approaches therefore suggest that low-ionisation lines probe denser gas. To understand the physical implications of this, we can look at galactic wind models. If we take for simplicity the spherical, momentum driven wind model of \citet{murray05}, the gas density is predicted to rapidly decline with radius ($\sim r^{-2}$ - $r^{-3}$), which is also reasonable considering that the wind is subject to geometrical dilution at larger distances from the galactic centre. 
The same model predicts that the wind is accelerating as an effect of radiation pressure pushing onto dust grains, with velocity increasing towards larger radii. We notice that accelerating winds seem to also be more consistent with observations of local starburst galaxies \citep{martin09} and high-redshift Lyman-break galaxies, as discussed in \citet{steidel10}. They also better reproduce the resolved profile of the MgII absorption line in star-forming galaxies at $z\sim1.4$ \citep{weiner09}. 
It is therefore clear that, according to this simple model, highly ionised gas, which traces more rarefied gas regions far from the centre, should also move faster, which is exactly what we observe in our sample of star-forming galaxies at $z>2$. We stress again that this model is a simplistic assumption and is by no means the best or the only model describing star-formation-driven winds. However, despite all the assumptions made, we must recognise that it provides an easy framework with which to interpret our observations. 

In conclusion, the most plausible physical explanation of the velocity difference observed between low- and high-ionisation lines relies on the fact that they come from slightly different spatial regions of the galaxies, at different distances from the centre. Therefore, they likely trace gas in different conditions of density and temperature, with low-ionisation gas probing the denser and colder star-forming cores, while higher ionised material is more widespread, optically thin, and likely coming from the outer shell of the expanding winds. 

\section{Summary and Conclusions}\label{conclusions}

In this paper, we present our study of the ISM kinematics of $330$ star-forming galaxies with a systemic redshift estimated through \CIII or \HeII detection with S/N $\geq 3$ in the redshift range between $2$ and $4.5$ ($z_{median}\simeq3.1$), and in the stellar mass range from $10^{8}$ to $10^{11}$ \msun. We derived the kinematic properties from a large variety of ISM absorption lines, which are compared to the systemic redshift inferred from \CIII or \HeII, and then related to different physical properties of the host galaxies. 
We summarise the main findings of this paper as follows.

\begin{enumerate} 

\item The ISM in typical star-forming galaxies at redshifts $2$ to $5$ ($z_{median}\simeq 3$) and with stellar masses of between $10^8$ and $10^{11}$ \msun is globally in outflow, with a median bulk velocity \vism of $\sim -60 \pm 10$ km/s for low-ionisation gas traced by \xSiIIa, \xCII, \xSiIIb, and \xAlII, and $200>$ \vism $>150$ km/s for higher ionisation gas traced by \SiIV and \AlIII. Individual galaxies can have ISM velocities spanning the range $\pm1000$ km/s, with higher ionisation gas reaching higher velocities on average and with a larger dispersion around the median value.
\item Our sample includes a fraction of galaxies ($34 \%$) with positive velocity shift obtained from the combined fit of low-ionisation lines, indicating the prevalence of inflowing material in those systems. This fraction is approximately double that found in similar works at lower redshift \citep[e.g.][]{steidel10}, suggesting a more important role of inflows at earlier epochs in the Universe. 
\item We do not find significant correlations between the bulk ISM velocity shift (\vism, \vismcomb, \vismSiIV, \vismAlIII) and the stellar mass \mass or SFR of the galaxies, while a marginally significant (i.e. at $2 \sigma$ level) correlation is found with the size (probed by the equivalent radius \rt), light concentration parameter \ct, and SFR surface density \sigmasfr. 
\item In spectral stacks, we also measure the maximum velocity \vmax, defined as the velocity at which $2\%$ of the absorption line flux accumulates. 
We find that \vmax in our sample has on average a value of $\sim -500$ and $\sim -600$ km/s for the low-ionisation and high-ionisation gas (traced by \SiIV), respectively. 
\item Compared to the bulk velocity, \vmax is more closely related to galaxy properties. Galaxies with a higher stellar mass, SFR, concentration, \sigmasfr, and more compact size launch outflows with higher maximum velocities  on average, even though the correlations remain weak, with a significance of $\simeq 2 \sigma$ for SFR-related quantities (i.e. SFR and \sigmasfr), and slightly higher significance for the morphology-related parameters. 
\item We detect a complex shape of the \SiIV absorption feature, which shows a main absorption component arising in the ISM and closer to the systemic redshift, and a fainter, additional absorption component likely originating in stellar winds, which is blueshifted from the main absorption peak by $\sim1000$ km/s. We find that BPASS models with binary populations, at odds with the BC03 and S99 models, are able to perfectly reproduce the blueshifted stellar wind absorption component of the line. This complex shape can be detected in both spectral stacks and some individual galaxies with higher S/N. 
\item Mergers show similar or slightly lower outflow velocities compared to non-merging systems, suggesting that inflows and internal turbulence induced by the interaction might dilute the outflow signatures coming from star-formation feedback.
\item Considering the $196$ galaxies with a net outflow velocity (i.e. a negative \vismcomb), we obtain average outflow velocities of $-146 \pm 10$ km/s for the low-ionisation gas and $-270 \pm 30$ km/s for the high ionisation component traced by \SiIV. From them we calculate, in a spherically symmetric thin shell approximation, the mass outflow rate $\dot{M}_{out}$ of the gas, which is comparable to the star formation of the galaxies ($\eta \simeq 1.3$). The values of $\dot{M}_{out}$ and \vism are also in favour of radiation-pressure-driven outflows. 
\item We derive a rough estimate of the escape velocity expected in our galaxies of $v_{esc}$ $=625$ km/s. This is higher than the average bulk and also maximum outflow velocity derived for our sample, even though for a small fraction of galaxies we can have  $\mid$\vismcomb$\mid$ and $\mid$\vismSiIV$\mid$ $>$ $v_{esc}$.
\end{enumerate}

In the near future, spatially resolved studies with IFU instruments, such as NIRSpec with JWST and ERIS at the VLT, will provide better insight into the different components of the ISM in high-redshift galaxies, and into the possible local relationships between kinematics and physical properties. Their higher spectral resolution will also allow the intrinsic profile of absorption lines to be characterised, even for individual galaxies. Similarly, ALMA can provide spatially resolved maps on the cold molecular gas, which would help to better assess the nature and final fate of the outflows, constraining their role on galaxy evolution.

\noindent

\begin{acknowledgements}
We thank the anonymous referee for constructive comments. AC acknowledge the support from grant PRIN MIUR 2017 20173ML3WW$\_$001. RA acknowledges support from ANID Fondecyt Regular 1202007. 
\end{acknowledgements}

\appendix


\section{Properties of individual ISM lines} 

\subsection{Estimating the bias in the systemic redshift determination from CIII]}\label{appendix0}

\begin{figure}[ht!]
    \centering
    \includegraphics[angle=0,width=1\linewidth,trim={0cm 0.cm 0cm 0.cm},clip]{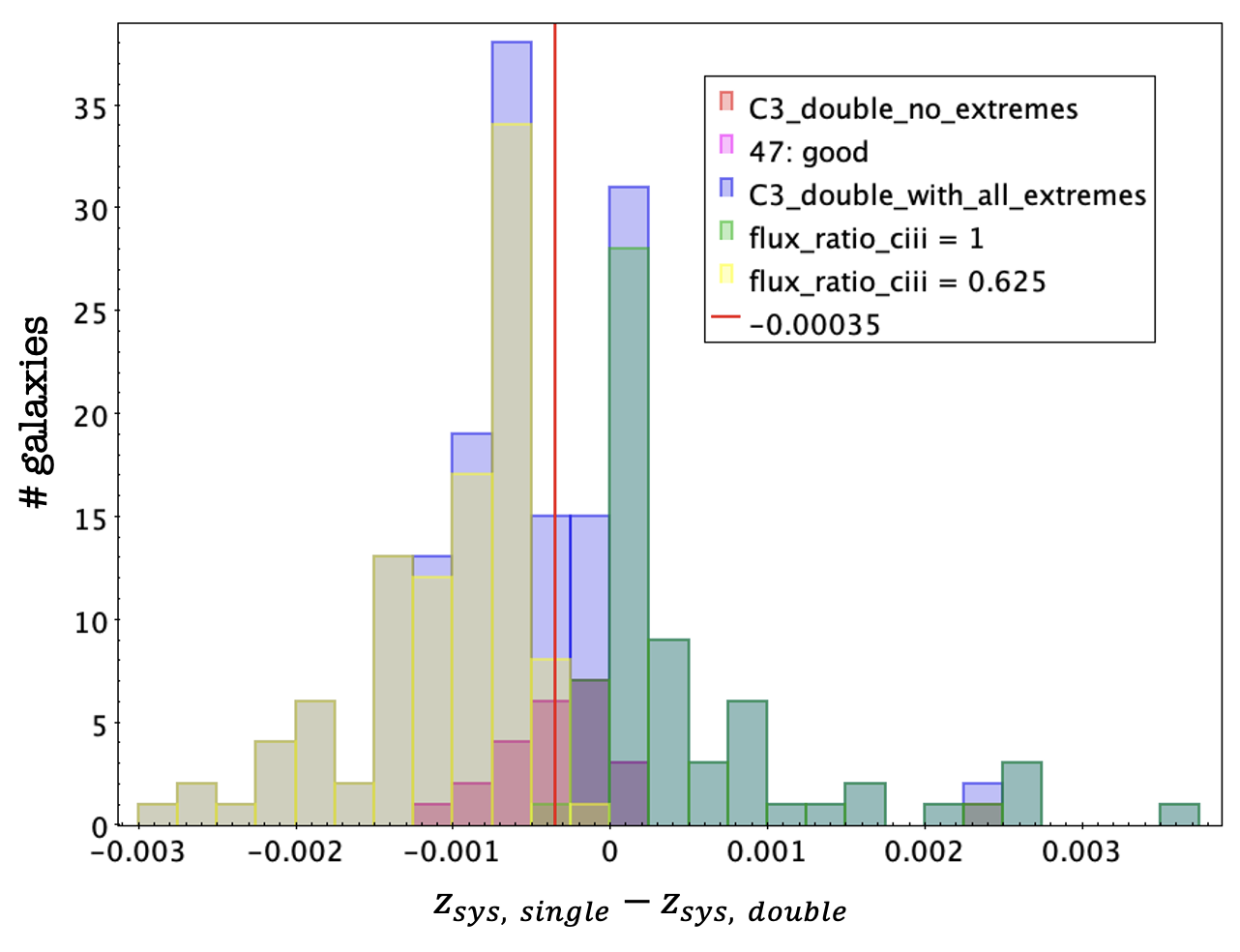}
    \includegraphics[angle=0,width=1\linewidth,trim={0cm 0.cm 0cm 0.cm},clip]{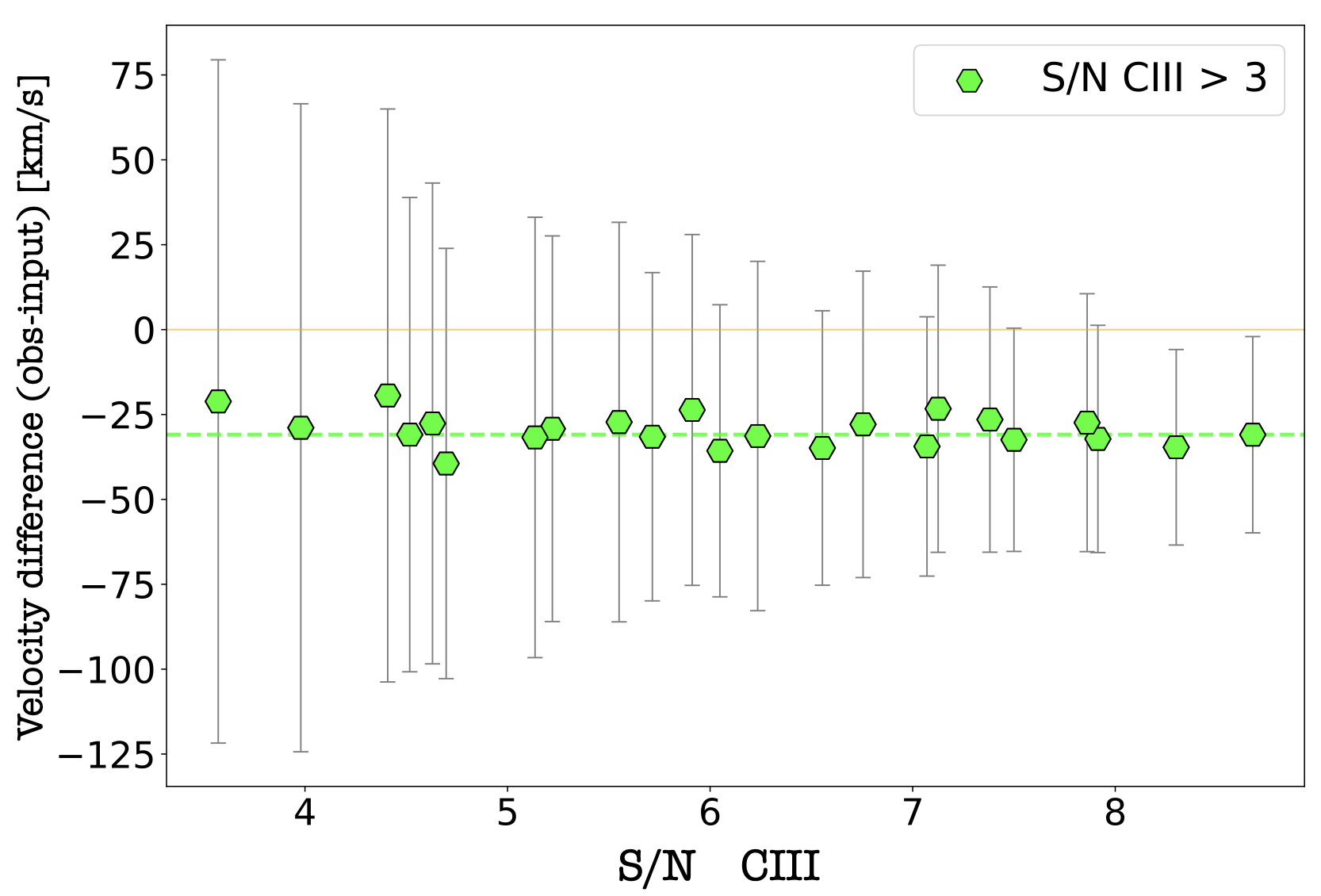}
    \vspace{-1mm} 
    \caption{\small \textit{Top panel:} Histogram distribution of the difference between the systemic redshift estimated from \CIII performing a single-Gaussian fit ($z_{sys,single}$) or a double-component fit ($z_{sys,double}$). The blue histogram is the sum of all galaxies with \CIII detected at S/N $\geq3$. The yellow and green histograms represent the subsets where the best-fit flux ratio between the two components of the double-Gaussian fit reaches one of the two extremes (respectively $1$ and $0.625$). The red histogram, from which we derive the correction of $26.25$ km/s for the systemic redshift estimated through a single-Gaussian fit, includes galaxies with a best-fit ratio within the allowed range. \textit{Bottom panel:} Velocity difference between the systemic redshift estimated from a single-Gaussian and a double-Gaussian fit of the \xCIII doublet, from Monte Carlo simulations, as a function of the S/N of the \CIII line. The horizontal green dashed line represents the median difference for the simulated dataset.}
    \label{zsystemic_test} 
\end{figure}

As described in Section \ref{systemic_redshift_estimation}, we determine the systemic redshift from the \CIII emission line. 
Given the spectral resolution of VANDELS and the S/N of the line, we cannot perform a good double-Gaussian fit for the majority of galaxies. 
A single-Gaussian fit at the central wavelength of the doublet yields generally the most reliable results. However, it is important to estimate possible biases on the true systemic redshift arising from this choice. 

To test this possibility, we first compare the histogram distribution of systemic redshifts derived using a single and a double Gaussian fitting, for the subset of galaxies with a good estimate of z$_\text{sys}$ from the second method. The distributions are shown in Fig. \ref{zsystemic_test}-\textit{top}, where we can see that the redshift from double Gaussian fitting is on average slightly lower than inferred through a single Gaussian fit, by a velocity shift of $26.25$ km/s on average. 
We further run Monte Carlo simulations to checker whether this effect is significant and is due to the different methodology adopted. We simulated synthetic double component \CIII lines and underlying continuum at redshift $3$ with the same wavelength sampling and resolution of VANDELS, varying the S/N of the continuum (from 2 to 10) and the intrinsic FWHM of the line (from $300$ to $500$ km/s), as found from VANDELS measurements. We then performed $400$ simulations by slightly varying the flux density at each wavelength according to the S/N chosen, and measuring each time the single Gaussian line peak with the same procedure adopted for the observations. We then analyse the median difference between the measured and intrinsic redshift as a function of the S/N and FWHM of the \CIII line. 
The outcome of this exercise is shown in Fig. \ref{zsystemic_test}-\textit{bottom}, where we note that the systemic redshift obtained from single-Gaussian fit is systematically lower than the true redshift by $\sim30$ km/s on average, and this offset does not depend on the S/N of the line or on its FWHM. 
Therefore, even though this offset is typically lower than the uncertainty of our ISM-shift values, we decided to apply a correction of $+30$ km/s  to our measurements of the systemic redshift 
systematically (i.e. $-30$ km/s on the derived ISM-shift of absorption lines) when it is determined from a single-Gaussian fitting.

\subsection{Comparison of the systemic redshifts estimated through CIII] and HeII}\label{appendix1}

\begin{figure}[t!]
    \centering
    \includegraphics[angle=0,width=1\linewidth,trim={0cm 0.cm 0cm 0.cm},clip]{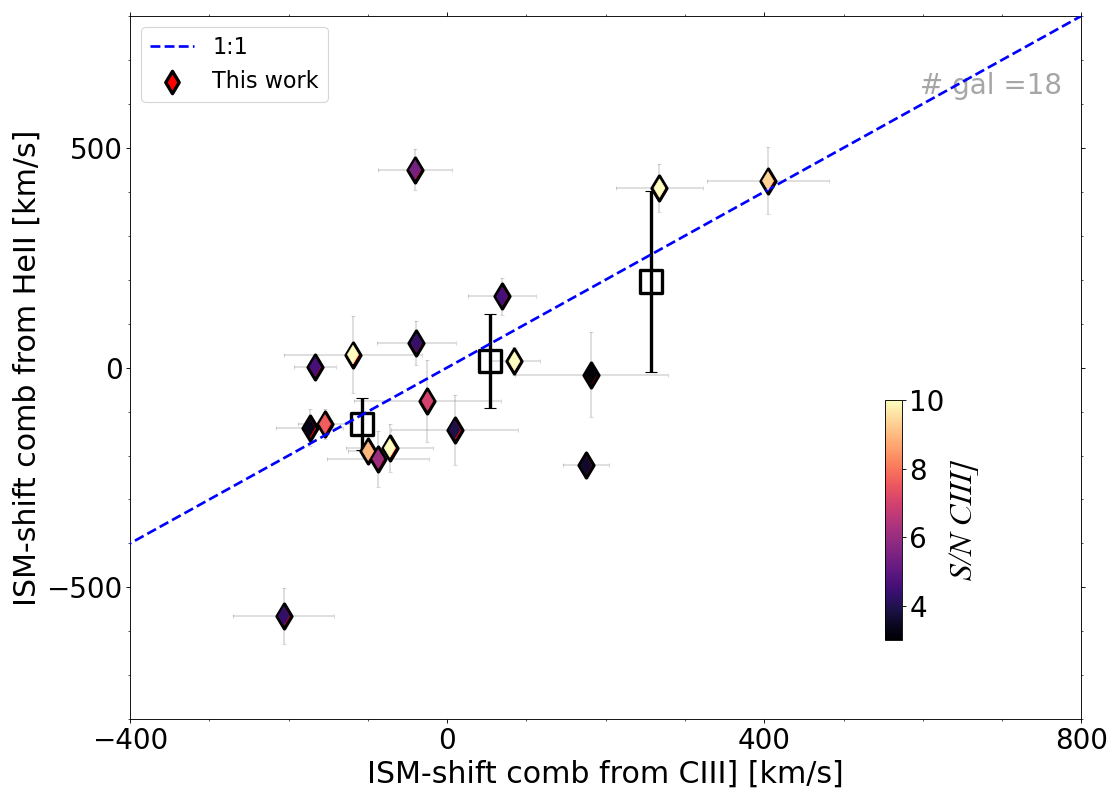}
    \vspace{-1mm} 
    \caption{\small Comparison between the combined ISM-shift (i.e. for lower ionisation lines) estimated assuming the \CIII as the systemic redshift indicator (x-axis) and the \HeII line (y-axis), for the sample where we obtain a reliable measurement of both lines with a S/N $\geq3$. The 1:1 relation is indicated with a blue dashed line, while the best-fit relation for our galaxies is indicated with black empty squares and black error bars. The data points are colour coded according to the S/N of the CIII] line.}
    \label{CIII_HeII_zsys_comparison} 
\end{figure}

In this section, we compare the velocity shift \vismcomb, representative of low-ionisation absorption lines and available for the largest number of galaxies, estimated assuming the systemic redshift derived from the \xCIII line and from \HeII, for a subset of $18$ galaxies where both emission lines are well fitted and detected with a S/N larger than $3$. We remind that for all these cases we adopt the value inferred from \CIII as our final estimate. 
The result of this comparison is shown in Fig. \ref{CIII_HeII_zsys_comparison}. We can see that, assuming the \HeII as systemic redshift indicator, would produce no systematic biases in the final value of \vismcomb, and the $1\sigma$ scatter of the relation is consistent with the typical uncertainties of individual ISM-shift measurements. 
A similar behaviour is also found when comparing the ISM shift of individual lines (including also \AlIII and \SiIV) estimated using \CIII or \HeII line for the systemic redshift. 

We know that the strong winds produced by Wolf-Rayet (WF) stars can broaden the \HeII line and induce an additional shift compared to the systemic redshift \citep{shirazi12}. However, as found in that paper, a significant contribution from these stars typically yields FWHM(\HeII) $> 1000$ km/s in galaxies, but these cases are excluded from our analysis, as explained in Section \ref{line_fitting}. We refer to \citet{saxena20} for a more complete treatment of the full sample of \HeII emitters in VANDELS. 

\subsection{Choosing the set of ISM lines for the combined fit}\label{appendix2}

\begin{figure}[ht!]
    \centering
    \includegraphics[angle=0,width=1\linewidth,trim={0cm 0.cm 0cm 0.cm},clip]{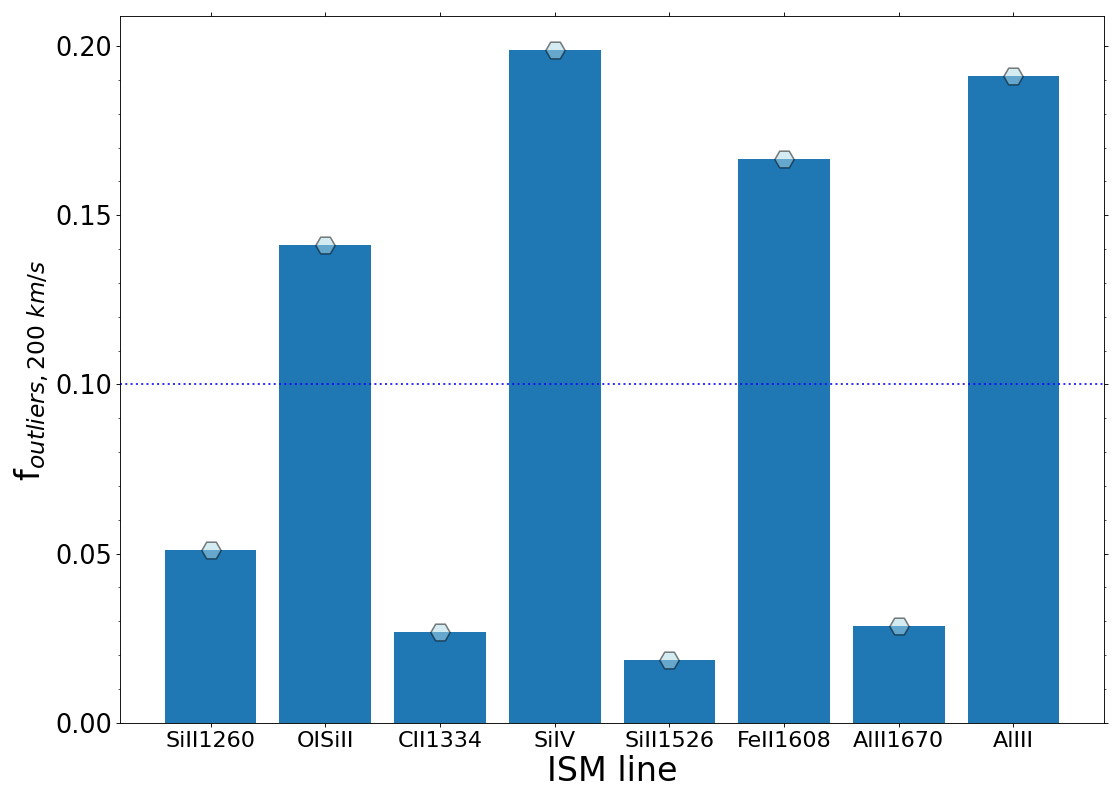}
    \vspace{-1mm} 
    \caption{\small  Fraction of galaxies for which the redshift measured from the single line fitting differs from the value obtained from a combined fit (including all the ISM lines detected in the spectrum) by more than $200$ km/s, for each ISM line analysed in this paper. The horizontal line represents the fraction ($10\%$) above which we have a significant number of outliers and the corresponding lines are not used in the combined fit.}
    \label{outliers_test} 
\end{figure}

\begin{figure*}[ht!]
    \centering
    \includegraphics[angle=0,width=0.95\linewidth,trim={0cm 10.cm 0cm 0.cm},clip]{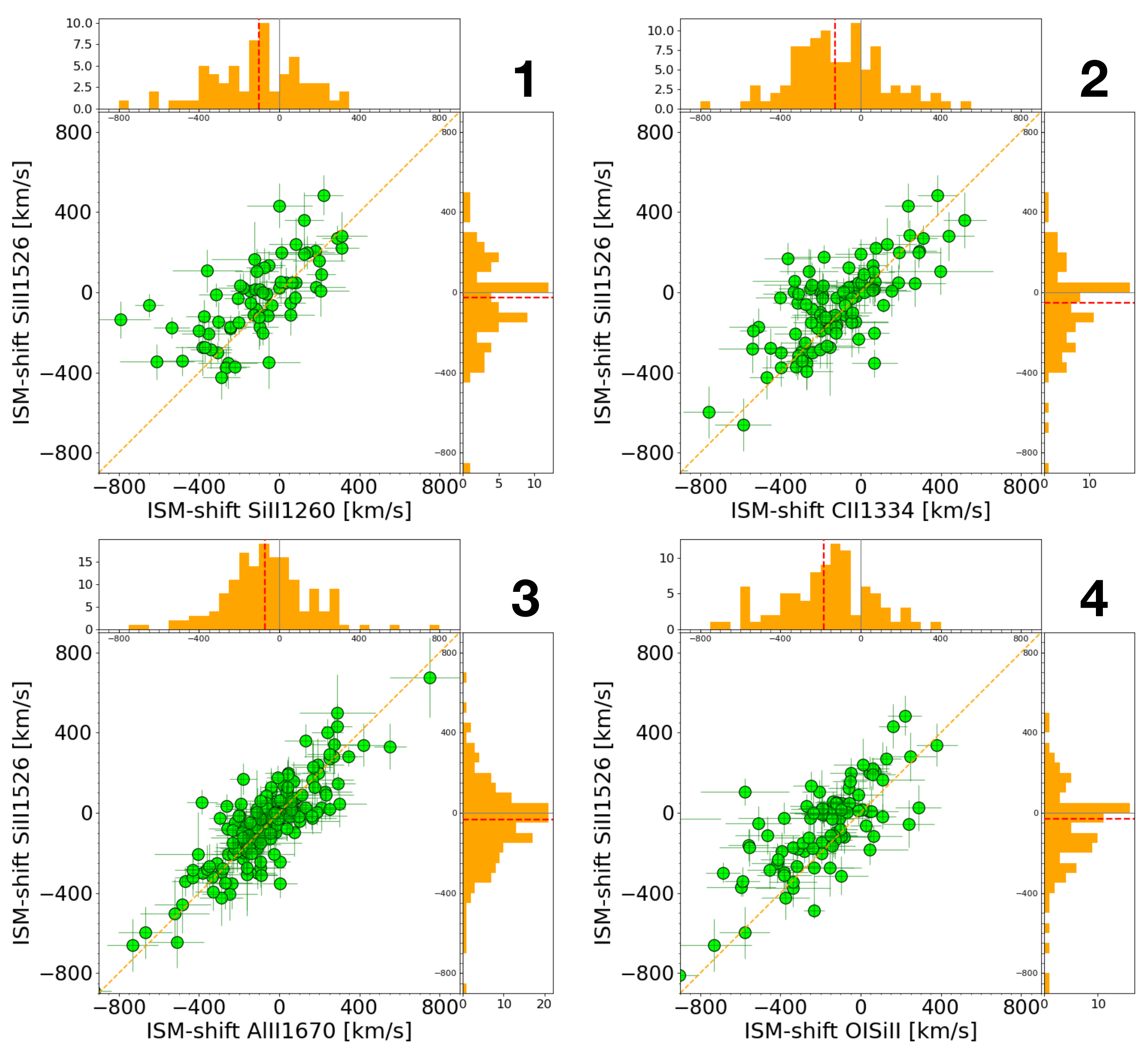}
    \vspace{-1mm} 
    \caption{\small Comparison of the ISM velocity shift (in Km/s, referred to the systemic redshift) of the \SiIV and \AlIII absorption lines for those galaxies in VANDELS where both features are detected with a S/N$>2$.} 
    \label{fig_app_1}
\end{figure*}

\begin{figure*}[ht!]
    \centering
    \includegraphics[angle=0,width=0.95\linewidth,trim={0cm 10.cm 0cm 0.cm},clip]{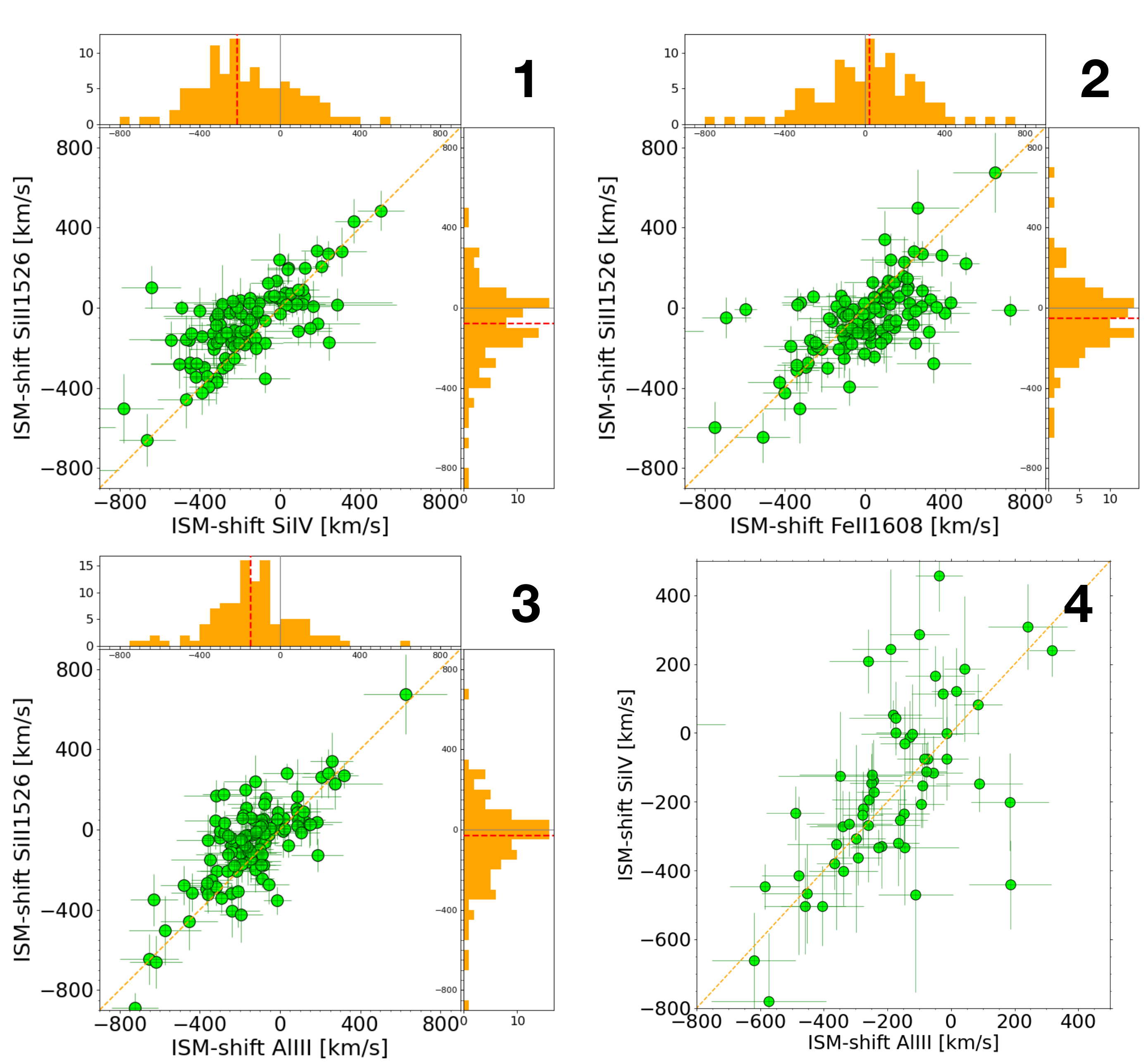}
    \vspace{-1mm} 
    \caption{\small Comparison of the ISM velocity shift (in Km/s, referred to the systemic redshift) of the \SiIV and \AlIII absorption lines for those galaxies in VANDELS where both features are detected with a S/N$>2$.} 
    \label{fig_app_2}
\end{figure*}

The large wavelength coverage of VANDELS spectra and the wide redshift range probed by the survey ($2<z<6.5$) allow us to study the ISM properties with multiple absorption lines in the UV regime. Most of these lines show similar properties in terms of velocity width and velocity shift with respect to the systemic redshift, hence it is reasonable to use them all together and perform a combined fit with a single redshift to increase the S/N of the final results. 
In order to decide which lines should be adopted in this combined fit, we first derive an average ISM redshift for each galaxy by fitting simultaneously all the ISM absorption lines that could be detected in the rest-frame range from $1216$ \r{A} to $2000$ \r{A}, namely \SiIIa $\lambda 1260$ \r{A}, \OISiII $\lambda\lambda 1302$-$1304$ \r{A}, \CII $\lambda 1334$ \r{A}, \SiIV $\lambda\lambda 1394$-$1403$ \r{A}, \SiIIb $\lambda 1526$ \r{A}, \FeII $\lambda 1608$ \r{A}, \AlII $\lambda 1670$ \r{A}, and \AlIII $\lambda\lambda 1855$-$1863$ \r{A}, as also listed in table \ref{tabella1}. Given that we are studying individual lines before the combined fit, we apply here a detection threshold of $3$ on the absorption lines, to increase the reliability of this test. 
We then compare these values to the redshift obtained for each individual absorption line used in the combined fit. Finally, for each line, we compute the fraction of galaxies where the corresponding redshift differs from the simultaneous fit value by more than $200$ km/s. A higher fraction for a specific line would imply that its properties are different from the average of the other ISM lines, and thus it should be excluded from the combined procedure. 

The outliers fraction f$_\text{outliers}$ for all the above ISM lines is presented in Fig. \ref{outliers_test}. We can clearly see a segregation in this histogram: while f$_\text{outliers}$ is very low (below $5\%$) for half of the lines, it increases by a factor of three to ten for the other half of the list. Defining a maximum limit for the outliers fraction of $10\%$ allows to isolate four lines, namely \SiIIa $\lambda1260$ \r{A}, \CII $\lambda1334$ \r{A}, \SiIIb $\lambda1526$ \r{A} and \AlII $\lambda1670$ \r{A}, that have likely similar properties and that we consider in this work for the combined fit. 

This choice is corroborated by a more detailed analysis of the entire set of ISM lines. Taking the SiII $\lambda1526$ \r{A} as our reference, as it is available for the largest number of galaxies in our sample, we plot the ISM-shift derived for this reference line against that obtained with all the other lines, taking subsets of galaxies for which the two absorption features are both detected. The results of this exercise are shown in the figures \ref{fig_app_1} and \ref{fig_app_2}. We can see that the four lines chosen for the combined fit are well aligned along the 1:1 relation with some scatter but no evident systematics (panels $1$, $2$, and $3$). 

As far as the other lines are concerned, the \OISiII $\lambda\lambda1302$-$1304$ \r{A} complex should be treated with caution. We performed a single Gaussian profile fit by considering the median of the doublet as the central wavelength (Fig. \ref{fig_app_1}- panel $4$), and also a double Gaussian fit with a free-varying ratio between the two components. In both cases, we obtain ISM redshifts that are consistent with the global ISM value for part of the sample, but significantly lower for a large subset of objects, which causes the fraction of outliers to rise up to $\sim14\%$, as shown in Fig. \ref{outliers_test}. This might reflect the presence of two different species that contribute to the absorption complex, whose relative importance and properties are difficult to model, especially for individual systems where we are limited in S/N. For these reasons, we do not consider this absorption complex in the simultaneous fit.

The \xFeII line has on average a positive velocity shift compared to the systemic redshift for this particular subset, as shown in Fig. \ref{fig_app_2}- panel 2, and we find a difference of $\sim +150$ km/s with respect to the ISM-shift of \xSiIIb. This might suggest that the FeII lines traces a gas component which is closer to the nucleus of the galaxy. As an alternative explanation, given that FeII is actually a blend of multiple hyperfine transitions of the same ionised element, there might be a non negligible contribution from transitions slightly longward of the main absorption at $1608.45$ \r{A} that is typically assumed also in other studies. 

Finally, the \SiIV $\lambda\lambda1394$-$1403$ \r{A} and \AlIII $\lambda\lambda1855$-$1863$ \r{A} doublets, which are representative of a gaseous component in a higher ionisation state, show on average more negative velocity shifts compared to the systemic redshift, indicative of higher velocity outflows by $\sim 150$ km/s, hence we should treat them separately (Fig. \ref{fig_app_2}- panels 1 and 3). Moreover, these two lines share very similar properties, as the \vism that we measure from them are well correlated and remarkably in agreement with the 1:1 relation, without evident systematic errors (Fig. \ref{fig_app_2}- panel 4). This suggests that they might be powered by the same mechanism and emitted in the same regions.

\end{document}